# Reflective Computing

Seven Dimensions from the Gaps in 134 Interactive Systems, 1996–2026

MD HASEEN AKHTAR

Department of Design, IIT Hyderabad, Telangana, India, haseen@des.iith.ac.in, ORCID 0000-0002-8019-0005

Interactive technology is increasingly built for speed, and generative AI now answers before a question is fully formed. Yet the capacity to pause, interpret and decide for oneself is central to wellbeing, learning and meaningful use. HCI has named paradigms for sensing emotion, changing behavior, calming attention and tracking the self. However, none of them takes human reflection, and the slowing down it requires, as its organizing aim. This article defines that paradigm, reflective computing, and derives it from what thirty years of designs lack. We analyzed 134 interactive systems from 17 ACM venues (1996–2026) in two phases. First, an abductive gap analysis analyzed every system against reflection theory, and seven recurring gaps defined seven dimensions, namely tempo agency, invitational form, interpretive openness, reflective depth, temporal span, protected idleness and ethical non-extraction. Second, two independent coders scored every system with a codebook (Krippendorff's alpha 0.67 to 0.84 for five dimensions; ICC 0.82 for the overall score). The system sets the pace in 51 designs, 74 support only momentary use, only 12 scaffold reflection beyond description, and 56 papers are silent on the inner states they collect. No system expresses all seven dimensions strongly, and only 20 meet the definition. Slow and experience-centered design comes closest, whereas persuasive technology and human-AI interaction score lowest. Factor and latent class analyses reveal two cultures of reflection, one working through form and the other through scaffolded thought, which only one system combines. Systems that let people set the pace score markedly higher, even with paradigm and form controlled (odds ratio 13.5). Hence, reflective computing is proposed as a paradigm whose aim is a human capacity rather than a system function, and which treats slowness as valuable only when people choose it.

*before we begin*_The lift doors close. There are four of us inside, and four screens light up at the same time. No one speaks. No one looks up. The ride takes seven seconds. I see this everywhere: on the stairs, in meetings, at coffee breaks, in every small gap between one thing and the next. In the last ten years, life has become much faster. We feel it, even if we cannot explain it. Content never stops coming, and everything is competing for our attention. For many of us, the phone has become the most addictive thing we own. We no longer pick it up and put it down. We live through it. Sometimes it feels as if we breathe through our phones. I struggle with this too, and so does almost everyone I know. Every minute brings another small reward, another alert, another reason to be somewhere else. It becomes harder to think deeply, harder to enjoy life, and harder to stay close to the people we love. Very few designs have truly helped people break this habit, and this is not an accident. Look at almost any new product today, and you will see that it is designed to catch our attention. For years, designers, psychologists and thinkers have told us to slow down, to let ourselves be bored and to put the phone away. We hear them, but we cannot do it. Distraction is no longer just a habit. It has become the way we live.

*where do we begin*_This journey started a year ago, at my desk. I was ideating on a new research direction, so I began finding patterns in design interventions that HCI and design research had produced over the last three decades, and I found something interesting. Only a small part of this work was truly about people, about how we connect with each other and with our own lives. Much more of it, often without meaning to, added to the fight for our attention. Each new design asked for a little more of our time. I understand the urge to create something, to contribute and to feel that our work matters. I have felt it too. However, as I was ideating, I became more interested in a different question. What if design did not always ask for more of people's time and attention, but sometimes helped them reclaim it? Then I looked up from my desk and out of the window. It is a busy morning on campus. Students ride their bicycles to the lecture hall, talking as they go. A group walks to the library, laughing and joking. The cleaning staff sweep the paths and call out to each other. This is the world we have made, and I always thought it was the world we wanted. Now I am not so sure. I imagine the same scene ten years from now, and I wonder if the phone was only the start. As AI moves into robots and daily services, many of those jobs may be done by machines. The jokes on the way to the library may become talk between a person and a robot. Robot carts may carry students to class, and no one will need a bicycle. The exercise would be gone, and so would the curious, unplanned talks of the morning ride. In their place, something would always be ready to pull the students' attention away.

*how do we begin*_This change has already started in small ways, and each step looks harmless, but it is not. I remember, as a student, waiting outside my professor's office. While I waited, I practiced my question, thought about what to discuss and felt eager to show my progress. Today, a student can ask a chatbot at midnight to complete a task and get an answer in seconds. The answer helps, but the waiting, the practice and the conversation that came after are gone. Instead of reading a paper slowly, getting lost and finding our way, we can read a short summary. But it was in getting lost that we often found what we did not know we were looking for. A delivery robot leaves our food at the door, and the few words we used to share with the delivery person are gone too. Apps choose what we watch, what we buy and even whom we meet, before we have time to wonder. AI companions are always there, always agree with us and never get tired. Slowly, they may make real friendships, which take patience and are never perfect, feel like too much work.

*should we begin*_I want to be honest about where I stand. I am not against technology. A chatbot that helps a student at midnight can make a real difference, and a robot that does dangerous and degrading work can be a true gift. My worry is more directed, and harder to see. With each new convenience, something else decides for us when we stop and when we move on, whether we struggle or are saved the effort, and whether we wonder or are simply told. Each convenience takes away a little effort, and with it, a little of the time we once used to notice, to question and to connect. We almost never choose this. It is chosen for us, one design at a time. Thus, this article asks a simple question. What would technology look like if it gave us back the time to pause, to wonder and to think for ourselves? In that kind of technology, we would set our own pace. We would find our own meaning. We would keep our quiet moments for ourselves and slowing down would be something we choose, not something chosen for us.

*let's begin*_

## 1 INTRODUCTION

Contemporary interactive technology is built to be fast. Feeds refresh without end, autoplay removes the decision to continue, and conversational AI answers before a question has been fully formed. The human capacity that this speed threatens most is one that Human Computer Interaction (HCI) has long valued, namely reflection. Dewey described reflective thought as the active, persistent and careful consideration of a belief or experience in the light of its grounds and consequences [36], and Schön showed that skilled practitioners reflect both during and after action [162]. Reflection, however, takes time, and dual-process accounts of cognition suggest that environments which reward speed displace it [90]. It is therefore imperative to ask how interactive systems can give people the time and the space to reflect.

Over the last three decades, HCI has organized much of its design work into named computing paradigms, each with its own aim for technology. Calm technology asked computing to inform without demanding attention [177], and affective computing asked systems to relate to, arise from or influence emotion [149]. Persuasive technology aimed to change attitudes and behavior [47], while positive computing set wellbeing and human potential as the goal [16]. Similarly, slow technology proposed designing for reflection and moments of mental rest [68], and personal informatics placed reflection within the cycles of self-tracking [116]. More recently, digital wellbeing has responded to compulsive use with tools for self-control [123, 132], and human-AI interaction now examines how people reason alongside intelligent systems [1, 169].

Reflection appears in many of these traditions, but only as a means, a stage or a side effect, and no paradigm makes it the organizing aim. Moreover, reviews have shown that HCI often invokes reflection without defining or evaluating it [7]. Frameworks for reflection have mostly described levels of thought [46] or conceptual dimensions [8], rather than the qualities that a designed system needs to invite, protect and deepen reflection. The closest relatives, reflective design [164], slow technology [68] and reflective informatics [8], each address part of this question, and Section 7.2 sets out how reflective computing differs from them. Thus, an important question remains open, and it is the question this article addresses. What would a computing paradigm look like if reflection were its aim?

This article proposes reflective computing as that missing paradigm and grounds it in the record of what HCI has built. Rather than defining the paradigm in the abstract, we analyze thirty years of designed interventions published at ACM venues and ask what they lack. The recurring gaps define the dimensions of reflective computing, and the dimensions are then coded by two independent coders and used to profile each system and paradigm. In this regard, four research questions guide the work. The first (RQ1) asks what recurring gaps prevent the designed interventions in ACM HCI venues from supporting reflection, and what dimensions of reflective computing those gaps define. The second (RQ2) asks whether the dimensions can be coded reliably by independent coders. The third (RQ3) asks from which computing paradigms these interventions were developed, and to what extent each intervention and paradigm expresses the dimensions. The fourth (RQ4) asks what each paradigm lacks to move toward reflective computing, and what design moves would close the gap.

The article makes four contributions. First, it gives a definition of reflective computing, with explicit membership conditions and exclusions, and seven dimensions, each grounded in a recurring gap across 134 systems. Second, it offers a codebook for the dimensions, reports its reliability across two independent coders, and releases the coded corpus with a reason for every score. Third, it presents a cross-paradigm profile that shows where reflective qualities are concentrated and where they are absent, including a statistically supported split between two cultures of reflection and a robust advantage for designs in which people set the pace. Fourth, it provides a paradigm-by-paradigm account of what is missing, with translation moves that carry the strengths of each paradigm toward reflective computing.

## 2 BACKGROUND AND RELATED WORK

### 2.1 Computing paradigms as design traditions

A computing paradigm, in the sense used here, is a named tradition that gives designers a shared aim, a vocabulary and a set of exemplars. Paradigms overlap, and many systems draw on more than one. Nonetheless, paradigms shape what designers notice and optimize. Table 1 lists the eleven paradigms that we used to classify the corpus.

Table 1: Computing paradigms used to classify the corpus.

| Code | Paradigm | Aim it sets for technology | Defining sources |
|---|---|---|---|
| CALM | Calm and ambient computing | Inform at the periphery without demanding attention | [152, 177] |
| AFF | Affective computing | Sense, express, communicate or regulate emotion | [149] |
| PERS | Persuasive technology | Change attitudes or behavior toward a target | [27, 47] |
| PI | Personal informatics | Collect and reflect on personal data | [41, 116] |
| POS | Positive computing and mental health | Support wellbeing, flourishing and mental health | [16, 147] |
| MIND | Contemplative and mindfulness computing | Hold attention gently on breath, body and present experience | [80, 170] |
| SLOW | Slow, ludic and experience-centered design | Make time perceptible, invite curiosity and open interpretation | [52, 68, 164] |
| DW | Digital wellbeing and self-control | Help people regulate their technology use | [123, 132] |
| CONN | Social connectedness computing | Support awareness and intimacy between people | [13, 42] |
| ATT | Attention and interruption management | Protect focus and manage notifications | [15, 186] |
| HAI | Human-AI interaction | Support people working and deciding with AI | [1, 169] |

### 2.2 Theoretical lenses on reflection

The analysis was sensitized by concepts from four bodies of work, namely classical accounts of reflection, frameworks developed within HCI, design traditions that treat form and pace as resources, and research on the conditions that support or undermine reflective thought. Each of these is described briefly below.

The first body of work concerns what reflection is. The idea that reflection begins when habitual action is interrupted comes from Dewey, who described reflective thought as the active, persistent and careful consideration of a belief or experience in the light of its grounds and consequences [36]. Schön extended this view to professional practice and showed that reflection can occur in action, while a task is under way, and on action, after it is complete [162]. Together, these accounts establish two points that matter for design. Reflection requires a pause, and that pause can be placed at different moments relative to the activity it concerns.

The second body of work concerns how HCI has described reflection. Fleck and Fitzpatrick proposed levels of reflection that run from description, through reflective description and dialogic reflection, to transformative and critical reflection [46]. Baumer identified breakdown, inquiry and transformation as conceptual dimensions of reflection [8], and Bentvelzen et al. derived four design resources from reflection research, namely temporal perspective, conversation, comparison and discovery [11]. These frameworks describe either the stages of reflective thought or the means of prompting it. However, none of them sets out the qualities that a designed system would need to invite, protect and deepen reflection.

The third body of work treats form, pace and openness as design resources. Slow technology argued that the form and pace of an artifact can make time perceptible and create moments of mental rest [68]. Furthermore, work on ambiguity and reflective design showed that leaving the meaning of a system open to interpretation can be a resource rather than a flaw [52, 164]. Together, these traditions moved attention from what a system tells people to how it invites them to think for themselves.

The fourth body of work concerns the conditions around reflection. Research on the sense of agency has shown that features people initiate support agency better than features a system pushes on them [120], which is consistent with self-determination theory and its view of autonomy as a basic psychological need [159]. Psychological research on boredom has described unfilled time as a signal and a resource rather than only a deficit [38]. In addition, dual-process accounts suggest that environments which reward speed displace deliberate thought [90], and work on dark patterns documents how interfaces exploit fast, automatic choice [64].

It is important to note that these concepts were used as lenses for reading the corpus, not as a predefined coding scheme. They helped the analysis notice what a system lacked, while the gaps themselves, and the dimensions they define, emerged from the systems (Section 3.3).

### 2.3 Prior reviews of reflection in HCI

Reflection has been reviewed in HCI before, but always within a particular boundary. Baumer et al. reviewed how reflection is used in interactive system design and found that the term is often invoked without being defined or evaluated [7]. Bentvelzen et al. derived design resources from reflection research [11]. Other reviews have examined single paradigms, such as mindfulness technology [170] and digital self-control tools [123, 132], and methodological work has called for more careful evaluation of behavior-change systems [102] and for more transparent literature reviews in HCI [167]. Nevertheless, these reviews leave three gaps that the present article addresses.

First, they stay within one paradigm. Each reads systems inside a single paradigm, or reads reflection as a concept, rather than comparing designed systems across paradigms. As a result, what one tradition has learned about reflection is rarely brought to bear on another. Second, they catalog what exists. They describe what systems contain rather than what systems collectively lack, yet a system's shortcomings for reflection become visible only when it is analyzed against a standard it was not built to meet. Third, they do not lead to a paradigm, since none of them uses its findings to define a paradigm in which reflection is the organizing aim.

## 3 METHOD

### 3.1 Review design

The review is a critical review in the sense of Grant and Booth's typology [63]. It goes beyond describing the literature to analyze it critically and make a conceptual contribution, which here is the set of dimensions of a computing paradigm. Its corpus is a purposive set of exemplary designs, as in a design-space analysis, and the resulting dimensions are then tested for coding reliability. The article is single-authored. "We" is used as the authorial voice, and "the author" refers specifically to the author's role as first coder and adjudicator, to distinguish it from the second coder. Because the corpus was not produced by an exhaustive database search, it does not report PRISMA (Preferred Reporting Items for Systematic Reviews and Meta-Analyses) counts for the exemplar set, and Section 7.10 discusses what this design can and cannot claim.

### 3.2 Corpus selection

We sought interactive systems designed as interventions in the experience, behavior, emotion, attention, wellbeing or understanding of people, published at an ACM venue between 1996 and 2026. Many of them are Research through Design artifacts, whose value lies partly in the questions they make tangible [185]. Pure performance tools, input techniques and sensing studies without a user-facing intervention were out of scope, with one exception, as a few affective sensing systems were retained because they define the lineage against which later affective designs reacted. In addition, a small number of included papers present a design practice, probe or intervention rather than a deployed artifact (for example, the time capsule study [148] and the 24-hour notification-free intervention [150]). These were coded on the practice or intervention they designed and are flagged in Appendix A.

Systems were identified from three sources, namely the author's knowledge of the field, the reference lists of paradigm-defining works and reviews [7, 11, 123, 132, 170], and backward citation chasing, that is, following the references of systems already included to find earlier systems they built on. The intention was to cover each paradigm across the full period, with emphasis on systems widely cited as exemplars. The corpus is therefore purposive, meaning that systems were chosen deliberately as representative examples rather than drawn at random. As a result, its paradigm counts describe the exemplar set rather than the distribution of ACM publications.

Each selected system was verified against an independent bibliographic source (ACM Digital Library, DOI resolver or publisher repository), confirming that it is an archival ACM publication whose title, authors, venue and year match the published record. The final corpus comprises 134 systems from 17 ACM venues (Table 2).

Table 2: Included systems by venue family.

| Venue family | Systems (n) |
|---|---|
| CHI (full papers, extended abstracts, CHI 98 summary) | 76 |
| UbiComp and IMWUT | 17 |
| CSCW and PACM HCI (CSCW) | 15 |
| DIS | 10 |
| UIST | 5 |
| IUI | 3 |
| MobileHCI | 2 |
| AVI, CHI PLAY, TEI, TOCHI, IDC, FAT* (one each) | 6 |
| **Total** | **134** |

### 3.3 Abductive gap analysis (Phase 1)

The first phase asked one question of each system. If this design were meant to invite, protect and deepen reflection, what would it lack? Each system was analyzed from its published description, and the answer was recorded as a short gap note. At the same time, we recorded the paradigm of each system, its form, its domain and who controls the pace of interaction.

The analysis was abductive rather than purely inductive. In simple terms, it did not start from a fixed theory, nor did it ignore theory. Instead, it moved back and forth between what the systems showed and existing ideas about reflection until the two fitted. In practice, it moved between the gap notes and the sensitizing concepts of Section 2.2, following reflexive thematic procedures [12]. Notes were clustered by the kind of absence they described. Clusters were then named, compared with the sensitizing concepts, split or merged, and re-checked against the notes until each described one distinct, recurring absence that could be stated as a design quality. This process produced seven gaps. Each gap was restated positively as a dimension of reflective computing, that is, the quality a design would need to close that gap. It is worth noting that social reflection did not emerge as a separate recurring gap. Supplement S3 documents each gap with its prevalence in the final consensus coding, the sensitizing concept it relates to, and the systems that show and close it.

### 3.4 Codebook development and profiling (Phase 2)

The second phase turned each dimension into a three-level rubric (0 absent, 1 partial, 2 strong) and then into a written codebook. A codebook is a set of written instructions that tells a coder, that is, a person reading a paper and assigning scores, how to decide which score a system deserves, so that different people reach the same decision for the same reasons. For each dimension, the codebook states the question the coder asks, a definition of each level, and decision rules for borderline cases (Table 4, with the full codebook in Supplement S2, which also gives an illustrative anchor example from the consensus coding for every level). Codebook v1.0 was frozen before any second coding began. A **proximity score**, the sum of the seven-dimension scores (0–14), summarizes how many reflective qualities a design expresses. The higher the score, the closer a system comes to reflective computing. However, it is not a measure of a design's quality on its own terms, since a persuasive system can be excellent at what it set out to do and still score low.

In addition, four further variables were coded for every system. **Pace** records who set the tempo of interaction (System, Shared or User), that is, whether the system decides when people interact with it, whether both sides do, or whether the person alone decides. **Form** records the form people interact with most (screen, tangible, ambient, wearable, conversational, immersive or hybrid), in other words the physical shape the technology takes in a person's life. **Reflection aims** records whether reflection by the user is a primary stated aim of the design (2), a secondary or instrumental aim (1), or not an aim (0), judged from the title, abstract and stated design goals. **D7 reporting** separates papers that are silent on what happens to users' data, emotions or reflections (coded NR, not reported) from papers that describe their data practice without making it a design commitment (coded 1). NR is scored as 1 in the proximity score so that silence is neither rewarded nor penalized and is analyzed separately (Section 5.9). Two descriptive variables were also recorded, namely the **domain**, the area of life a system serves (for example health, work or the home), and the **period**, the five-year span in which it was published.

### 3.5 Independent coding and adjudication

Every system was coded independently by a second coder, a human HCI researcher with five years of experience in qualitative and design analysis, including prior experience of coding research literature against a codebook, and familiar with research through design, slow technology, personal informatics, affective computing, persuasive technology and human-AI interaction. Because the second coder was already trained in this kind of analysis, they coded directly from the written codebook. The second coder took no part in Phase 1 and was not involved in deriving the dimensions, while the author was the first coder. Coding was carried out in multiple sessions over six months. To keep the coding blind, the second coder received only the frozen codebook and the list of systems with their bibliographic references, in shuffled order and split across seven batches. Thus, they did not receive the author's scores, the paradigm classification, the gap notes or the manuscript, and never saw the author's scores before the agreement round.

For each system, the second coder located and analyzed the published paper. In the first round, they worked from the full text for 85 systems, from the abstract for 20, from secondary descriptions by the same authors (theses, project pages, later articles on the same system) for 23, and from their prior knowledge of the paper for 6. After adjudication, the author supplied the full text of every paper that had not been analyzed in full, and the second coder re-coded each from it (see the full-text re-check below). Hence, the final dataset rests on the full text for 134 of 134 systems. Each score carries a one-line reason that points to evidence, together with a confidence rating.

Reliability here means the extent to which two people who apply the same codebook independently arrive at the same scores. It matters because, without it, the results would reflect one person's judgment rather than the codebook. Agreement before adjudication was computed for each dimension in four complementary ways. Percent exact agreement is the simplest, being the share of systems on which both coders gave the same score. However, two coders would agree on some systems by chance alone. Hence, we also report three chance-corrected measures, which run from 0 (no better than chance) to 1 (perfect agreement). Cohen's κ [23] is the most widely used of these. We use quadratic weights, so that a disagreement between 0 and 2 counts as more serious than one between 0 and 1. Krippendorff's α for ordinal data [110] is the standard measure in content analysis and is the one we use to judge whether a dimension is reliable. Gwet's AC2 with quadratic weights [67] is reported because κ and α are depressed when one category dominates, as on D7, where most systems sit at a single level [44]. This is known as the prevalence paradox, in which two coders agree almost always and yet receive a low score. Agreement on the proximity score is reported as the intraclass correlation ICC (2,1), which measures how closely two sets of numerical scores match. Krippendorff's thresholds were used to interpret α, with values of ≥ .80 taken as reliable and 0.667–0.80 as acceptable for tentative conclusions [110].

After the agreement round, four systematic disagreements were traced to decisions that the codebook had left open (Section 5.1). The author resolved each as a codebook amendment (v1.1, Table 4), without reference to individual systems' scores. Every remaining disagreement (309 cells in 126 systems) was then adjudicated, that is, settled by examining the evidence for both scores and deciding which one the codebook supports. For each cell, the coders prepared an adjudication note that set out both scores, the second coder's reason, the relevant codebook rule and the paper's evidence, re-reading the paper where the cell turned on a detail, and proposed a final score. The author reviewed every proposal and made the final decision. Therefore, the final scores are adjudicated scores, referred to below as consensus scores. The full adjudication log, including the cells that rest on abstracts or secondary sources (70 cells), is archived with the corpus.

In the full-text re-check, the 49 papers that the second coder had first coded from abstracts (20), secondary descriptions (23) or prior knowledge (6) were re-coded from the full text, using only the paper and codebook v1.1, without access to the author's scores or the provisional consensus, that is, the agreed scores reached at adjudication before this re-check. However, because the second coder had seen both coders' scores for disputed cells during adjudication, this re-coding was independent of the record but not blind. We therefore treat it as a check on source quality, not as a second agreement round, and all reliability estimates for independent coding come from the first round (Table 5). Where a full-text code differed from the provisional consensus, the change was listed with its evidence, and the author decided the final score. Every proposed change and decision is recorded in the full-text re-check log (Supplement S1).

Finally, alongside the seven dimensions, the second coder recorded for every system anything it lacked for supporting reflection that D1–D7 did not capture (an open "other gap" code). Because the second coder was blind to the derivation of the dimensions, these notes provide a partial test of whether the dimensions miss recurring absences (Section 5.10).

### 3.6 Statistical analysis

This section explains, in plain terms, how the coded scores were analyzed and why each method was chosen. Each paragraph below opens with the question it answers, so that readers who are less familiar with statistics can follow the logic without the technical detail, and the final paragraph lists the technical settings for readers who wish to reproduce the analyses. The analyses address four questions. First, do some dimensions tend to appear together? Second, do the dimensions form distinct groups, and do the systems form distinct types? Third, do the eleven paradigms differ? Fourth, is who sets the pace linked to the other reflective qualities?

***Why are special methods needed?*** Each dimension is scored 0, 1 or 2. These scores are ordinal, that is, they are ordered categories rather than true measurements. A score of 2 is higher than a score of 1, but it is not necessarily twice as much of a quality, in the same way that a gold medal ranks above a silver one without being worth exactly twice as much. Many familiar statistics, such as comparing averages with a t-test, assume true measurements. We therefore used methods designed for ranked or ordered data throughout. In addition, every estimate is reported with a 95% confidence interval (CI), which is the range within which the true value is likely to lie. A narrow interval means that the estimate is precise, whereas a wide interval means that the data are compatible with a broad range of values and that the result should be analyzed with caution. Where possible, the intervals were obtained by bootstrapping. Bootstrapping treats the 134 systems as a stand-in for the wider population, draws many new samples of 134 from them at random (allowing the same system to be drawn more than once), recalculates the result for each sample, and reads the uncertainty from how much the results vary. It makes no assumption about the shape of the data, which suits a small and uneven sample such as this one. Throughout the article, we rely mainly on the size of an effect and its confidence interval rather than on p-values alone, because the precision of an estimate is as informative as whether it passes a threshold.

***Question 1. Do some dimensions tend to appear together?*** If designers who make a system strong on one quality also tend to make it strong on another, the two dimensions are associated. We measured this with Spearman correlations (ρ). A Spearman correlation ranks the systems on each dimension and asks how far the two rankings agree. It runs from −1 to 1. A value of 1 means that systems ranked high on one dimension are always ranked high on the other, 0 means no relationship, and −1 means that a high rank on one always goes with a low rank on the other. As a rough guide, values around 0.1 are usually analyzed as weak, around 0.3 as moderate, and around 0.5 or above as strong. Spearman correlation was chosen because it uses ranks rather than raw values and therefore suits ordered scores. Each correlation is reported with a bootstrap 95% CI.

***Question 2. Do the dimensions form distinct groups, and do systems form distinct types*?** A correlation looks at only two dimensions at a time. To see the overall structure, we used two complementary methods that look at all the dimensions together. The first, exploratory factor analysis, asks whether the correlations among the dimensions can be explained by a smaller number of hidden factors. A factor can be thought of as an underlying theme, such as attention to form, that several dimensions share. For each dimension, the analysis gives a loading on each factor, which shows how strongly the dimension belongs to that factor. A loading near 1 means that it belongs strongly, and a loading near 0 means that it does not. Because factor analysis works with correlations between continuous qualities, it was run on polychoric correlations [144]. These estimate the correlation that would hold if each 0, 1 or 2 score were a coarse reading of an underlying continuous quality. To decide how many factors to keep, we used parallel analysis [81]. Parallel analysis compares each factor with factors obtained from random data of the same size and keeps a factor only if it explains more than chance would, which protects against reading structure into noise. The second method, latent class analysis, looks at the same data from the opposite direction. Instead of grouping dimensions, it groups systems into a small number of classes whose members have similar profiles of scores, much as one might sort students into groups by their pattern of grades across subjects. Models with 1–5 classes were compared using the Bayesian information criterion (BIC) [24]. The BIC rewards a model for fitting the data well but penalizes it for adding more classes. Hence, the preferred model, the one with the lowest BIC, is the simplest model that still describes the data well. If both methods point to the same groups, the finding does not depend on one technique. D7 was left out of both analyses because almost all systems share one level, so it carries too little variation to group.

***Question 3. Do the eleven paradigms differ?*** For each paradigm, we report the mean score with a bootstrap 95% CI [39]. Paradigms with six or fewer systems are flagged, because a mean based on so few systems is less certain. To test whether the paradigms differ overall, we used the Kruskal–Wallis test. This test ranks all 134 systems together and asks whether some paradigms sit higher or lower in the ranking than would be expected by chance. Its result is reported as H, where a larger H means larger differences between the groups, together with a p-value. The p-value is the probability of seeing differences at least this large if the paradigms did not really differ, so a small p-value (conventionally below .05) suggests that the differences are real. The Kruskal–Wallis test was chosen instead of the more familiar analysis of variance (ANOVA) because it compares ranks rather than means. It therefore suits ordered scores and remains reliable when, as here, the groups are small and unequal in size.

***Question 4. Is who sets the pace linked to the other reflective qualities?*** Here the question is whether user-paced systems score higher than system-paced ones. A simple comparison of averages could mislead, because pace is entangled with other characteristics. For example, slow designs tend to be tangible objects, and persuasive systems tend to be system-paced, so an apparent effect of pace might really be an effect of paradigm or form. We therefore used proportional-odds ordinal regression [127]. Like other regression models, it estimates the link between pace and the score while holding other variables constant. Unlike ordinary regression, it treats the outcome as an ordered score rather than a measurement. Its result is an odds ratio (OR). An OR of 1 means no association, and an OR above 1 means that systems in one group are more likely to have higher scores than those in the reference group. For instance, an OR of 2 means twice the odds of scoring higher. The outcome was the proximity score without D1, because D1 itself measures who controls the pace, and including it would make pace predict its own dimension. Paradigm and form were then added as controls one at a time, so that readers can see whether the link survives each adjustment. Two further checks were made. First, the rubric scores notifications, prompts and nudges as 0 on protected idleness (D6), and such prompts are usually sent by the system. A system-paced design can therefore lose points on D6 partly by definition, so the models were repeated with D6 also removed from the outcome. Second, the link was estimated for each dimension separately, to see where it lies. Testing several dimensions at once raises the chance that one of them appears significant by luck alone. We therefore adjusted the p-values with the Holm correction, which makes the threshold for significance stricter for each additional test.

***Data visualization.*** Co-occurrence of strong scores across dimensions is shown with an UpSet plot [115] (Figure 7), and the paradigm whose systems come closest to the strong level on each dimension is summarized in a dumbbell chart with a target marker (Figure 8). Each figure's chart type was chosen for the question it answers, and is named, with the reason for the choice, where the figure is first discussed. Throughout, we favor position along a common scale, which people read more accurately than angle, area or color [22], and use area or color only where a figure must show a pattern across a whole matrix rather than exact values.

***Technical settings.*** Confidence intervals for correlations used 2,000 bootstrap resamples, and those for paradigm means used 10,000 percentile bootstrap resamples. The factor analysis used minimum-residual extraction with oblimin rotation, which allows the factors to correlate with each other. The latent class analysis used categorical indicators. All analyses were run in Python (pandas, SciPy, statsmodels, factor_analyzer, StepMix) with a fixed random seed, so that every confidence interval can be reproduced exactly, and the scripts are archived with the data. Means are reported to one decimal place, rounded half to even, in both the text and the figures.

## 4 SEVEN GAPS THAT DEFINE REFLECTIVE COMPUTING (PHASE 1)

A gap (in the sense used here) is a quality that a system would need in order to support reflection but that it does not have, such as letting the person control the pace or leaving the meaning of an experience to them. Seven such gaps recur across the 134 systems. Each gap is described below with the evidence for it, counted on the final consensus scores, the systems that best show what closing it looks like, and the dimension it defines. Table 3 summarizes the result.

Table 3: Seven gaps across the corpus (consensus scores) and the dimensions of reflective computing they define.

| Gap observed across designed interventions | Systems (of 134) | Dimension defined |
|---|---|---|
| The system or designer sets the pace, and the person cannot slow, pause or decline | 51 system-paced; 34 with no pace control | D1 Tempo agency |
| Form is standard or urgency-oriented; nothing signals that pausing is welcome | 62 absent | D2 Invitational form |
| The system issues verdicts, scores or prescriptions, leaving no room for personal meaning | 53 absent | D3 Interpretive openness |
| Reflection is not targeted, or stops at recall and description | 58 absent; 64 stops at description; 12 scaffolds beyond | D4 Reflective depth |
| Use is momentary, with no support for looking back or ahead over time | 74 absent; 11 designed for months or years | D5 Temporal span |
| Idle moments are filled, or simply left unprotected | 32 fill idle time; 84 neither fill nor protect; 18 protect | D6 Protected idleness |
| Pauses, emotions or reflections are covertly influenced or exposed, or their use goes unreported | 13 covert or exposing; 56 silents; 2 committed | D7 Ethical non-extraction |

### 4.1 Pace imposed by the system defines tempo agency (D1)

In 51 systems (38%), the system or designer initiates and times essentially all interaction, and in 34 the person has no meaningful control over pace at all. The gap is most pronounced in persuasive technology, where the system sets the pace in 11 of 20 systems, and in affective computing (10 of 20). It also runs through slow technology. For example, Photobox prints photographs at intervals the household cannot change [138, 139], and the system or designer sets the pace in 7 of 19 slow designs. Similarly, AI decision aids that slow people down do so by imposing a pause, as cognitive forcing does [14]. Conversely, where people do control the pace, scores on the other dimensions are higher too. Excluding D1 itself, user-paced systems average 5.7, shared-pace systems 4.8 and system-paced systems 3.8 (Section 5.7).

Systems that close this gap give people the choice, such as personal goals in MyTime [74], user-switchable interface modes in SwitchTube [121], customizable tracking in Trackly [4], and breath-driven pacing in Sonic Cradle [175].

***D1 Tempo Agency.*** *The person, not the system, holds final control over the pace of interaction: they can slow down, speed up, pause, resume, or decline the system's invitations.*

### 4.2 Urgent or standard form defines invitational form (D2)

In 62 systems, nothing in the form of the technology signals that pausing is welcome. Screens dominate the corpus (56 systems), and 44 of them score 0 on form. Personal informatics (14 of 17), digital wellbeing (13 of 13) and human-AI interaction (12 of 12) are almost entirely without it. As a result, reflection in these systems is asked of people through the same interfaces that demand speed elsewhere.

The strongest counterexamples are physical, such as the slow aerial journey of the Drift Table [53], the lingering traces of the History Tablecloth [54], the slowly turning disc of Olly [141, 142], the crafted personhood jewelry of Wallace et al. [176] and the Spheres of Wellbeing [172]. In these designs, weight, pace and material quality tell people that taking time is the point.

***D2 Invitational Form.*** *The physical and digital form of the technology, including its material, motion, sound, rhythm and layout, signals that pausing is welcome and invites the person to stay.*

### 4.3 Verdicts and prescriptions define interpretive openness (D3)

In 53 systems, the system tells people what their experience means, in the form of a score, a goal met or missed, a recommended action, an emotional label or an AI answer. Persuasive technology (13 of 20 at 0), digital wellbeing (9 of 13) and human-AI interaction (8 of 12) are the main sources. Furthermore, some affective systems go further, reporting emotion as a measured quantity rather than something the person interprets [72, 119].

Systems that close this gap offer material rather than meaning, such as ambiguous ambient objects [55, 66], open channels between people [13, 18], sensed emotion returned as ambiguous expression [83, 84, 117], and questions in place of answers [34].

***D3 Interpretive Openness.*** *The system offers material, juxtapositions and questions rather than verdicts, and leaves the meaning of an experience to the person.*

### 4.4 Reflection that stops at description defines reflective depth (D4)

In all, 58 systems target no reflection at all, and a further 64 support only recall or description. That is, they show people what happened but do not help them ask why, consider alternatives or reframe. Only 12 systems scaffold reflection beyond description, through social scaffolding, explanation prompts, conversational follow-ups, self-experiments, comparison and question framing: MAHI [125], Echo [85], Visual cuts [40], Somaesthetic appreciation design [79], Self-experimentation approaches [112], TummyTrials [91], WorkAnalytics [129], Peer support chats [137], Reflection Companion [107], Robota [106], Question-framed explanations [34] and MindfulDiary [100].

***D4 Reflective Depth.*** *The design scaffolds the movement from describing an experience toward questioning, reframing and transforming one's understanding of it.*

### 4.5 Momentary use defines temporal span (D5)

In total, 74 systems, 55% of the corpus, support only momentary use. Nine of 11 mindfulness systems and 15 of 20 affective systems end when the session ends. Thus, reflection on meaningful change, which unfolds over weeks, months and years, is rarely supported. Only 11 systems are designed for revisiting or anticipation over long periods, most of them memory-oriented slow technologies and long-term records: Baby Steps [94], Time capsule [148], Photobox [138], Echo [85], Health Mashups [10], Personhood jewelry [176], Photobox field study [139], Inner Garden [158], Olly and Slow Game [141], Chronoscope [19] and Olly field study [142].

***D5 Temporal Span.*** *The technology supports reflection across meaningful periods of time, through revisiting, anticipation, long-term presence and graceful lapse and return.*

### 4.6 Unprotected idle moments define protected idleness (D6)

In 32 systems, idle moments are actively filled, or attention is demanded through notifications, prompts, timed nudges or feeds. Unlock journaling even uses the idle moment of unlocking a phone for data entry [183]. Persuasive technology (8 of 20) and personal informatics (9 of 17) are the main sources. The larger gap, however, is the absence of protection. In all, 84 systems neither fill nor shape idle time, and only 18 deliberately preserve or shape quiet. These include slow artifacts that do little most of the time [53, 138, 139, 141, 142], counterfunctional things [151], a workplace email cut-off [126], contemplative environments [79, 153, 158, 175], Mirror Ritual, which activates only when the person approaches [156], and digital-wellbeing designs that remove the newsfeed, autoplay or notifications [76, 104, 105, 121, 124, 150].

***D6 Protected Idleness.*** *The design preserves and shapes unfilled moments, so that idle time can support rest, mind-wandering and reflection rather than being filled automatically.*

### 4.7 Covert influence, exposure and silence define ethical non-extraction (D7)

Thirteen systems score 0 on this dimension: Second Messenger [37], Waterbot [3], Playful Bottle [20], Fit4life [154], BinCam [171], Mood Meter [72], EmotionCheck [28], Hint (color-changing shirt) [83], BrightBeat [62], EngageMeter [69], Let's FOCUS [97], Voice self-perception [29] and Opinionated LLM co-writing [89]. These systems covertly manipulate, surveil or expose people's inner states or behavior, through false physiological feedback [28, 62], covert voice manipulation [29], language models that shift users' opinions without their awareness [89], engagement sensing of audiences for presenters [69], smile counting in public spaces [72], public projection or sharing of individual behavior [3, 20, 37, 97, 171], and a garment that displays arousal without the wearer's control [83]. Fit4life embodies these failings on purpose as a critical design [154]. A further 56 papers say nothing about what happens to the emotions, reflections or behavioral data their systems collect, and 63 describe a data practice without making it a design commitment. Only two systems make the principle that reflection serves the person an explicit, embodied design commitment, namely Casablanca [73] and Spheres of Wellbeing [172]. As sensing and AI grow more capable, this gap widens.

***D7 Ethical Non-extraction.*** *A person's pauses, emotions and reflections serve that person: they are private by default, transparent in use, and never converted covertly into engagement, persuasion or data for others.*

### 4.8 Defining reflective computing

The seven gaps describe what HCI's designed interventions lack for reflection, whereas the seven dimensions describe what a reflective technology would need (Figure 1). Considering this, we define the seven dimensions and the paradigm as follows.

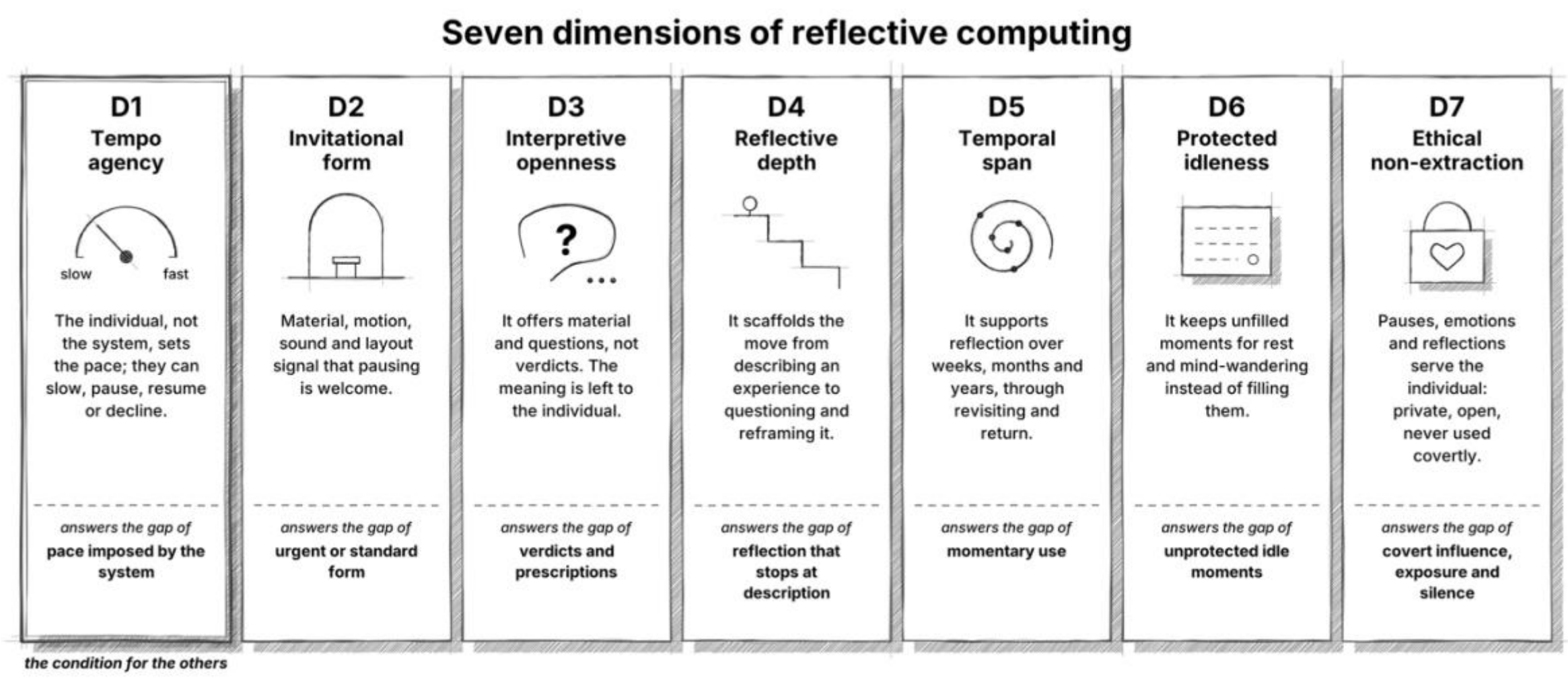

Figure 1: The seven dimensions of reflective computing. Each dimension is defined by the recurring gap it answers. Tempo agency (D1) is the condition under which the other dimensions become reflective (Author's own).

***Reflective computing*** is the design of interactive systems whose primary aim is to invite, protect and deepen a person's *own* reflection on their experience, beliefs or actions, at a pace the person controls. A system is an instance of reflective computing when (a) the user's reflection is the primary stated design aim, and (b) the design gives the person control over whether and when to reflect ($D1 \geq 1$), leaves the meaning of their experience to them ($D3 \geq 1$), and does not covertly use what they reveal ($D7 \neq 0$). The seven dimensions describe *how fully* a system realizes the paradigm, whereas these conditions describe *whether* it is an instance of it. Through this test, 20 of the 134 systems are instances of reflective computing, and the remainder are neighbors whose gaps define it.

A broader formulation, computing that "relates to, arises from, or deliberately influences" reflection, would echo Picard's definition of affective computing [149]. That breadth suits emotion, which almost every system touches. However, it does not suit reflection, since almost everything can be said to relate to reflection. Reflective computing therefore excludes seven kinds of system.

The first is machine reflection, that is, language-model agents that critique or revise their own outputs ("self-reflection"), and reflective or introspective programming in the software engineering sense. The second is reflection as a means to someone else's target, where prompts ask users to reflect so that they comply with a behavior, goal or recommendation chosen by the designer. Persuasive systems can move toward reflective computing (Section 6), but they are not instances of it while the target is fixed. The third is imposed slowing for task accuracy, such as cognitive forcing or friction that improves performance or reduces over-reliance, when the pause is not the individuals to take and reflection is not the aim. The fourth is sensing or modeling inner states without returning them to the person, as in affect detection, stress sensing or engagement metering.

The fifth is calm without sense-making, that is, breathing pacers or ambient displays that lower arousal or demand but do not invite people to consider their experience. Such systems may create the conditions for reflection (D2, D6) and still fall outside the paradigm. The sixth is reflection by designers or audiences rather than users. For instance, critical design such as Fit4life provokes reflection in the field, not in its users. The seventh is rumination and compulsive self-monitoring. Repetitive, self-critical or anxious review is not reflection, and a system that encourages it falls outside the paradigm even if it scores well on depth or temporal span.

Two features distinguish reflective computing from its neighbors. First, its organizing aim is a human capacity, reflection, rather than a system function such as sensing, persuading or calming. Second, it places tempo agency at its center, since slowness is valuable only when people can choose it. Hence, slow design, the shaping of tempo, form and idle time, and scaffolded thought, the shaping of prompts, questions and records, are its two principal means (Section 5.6).

### 4.9 From dimensions to a coding rubric

Table 4 states the rubric used to score each dimension, together with the principal decision rules of codebook v1.1. The full codebook is given in Supplement S2, which also provides an anchor example from the consensus coding for every level.

Table 4: Rubric and principal decision rules of codebook v1.1 (amendments A1–A4 adopted after the first agreement round; Pace: A4).

| Dimension | 0 Absent | 1 Partial | 2 Strong | Key decision rules (v1.1) |
|---|---|---|---|---|
| D1 Tempo agency | Pace fixed by system or designer | Person can adjust some aspects or opt out; or pace genuinely shared | Person sets or declines the pace throughout | Being able to switch the system off scores at most 1; dismissible fixed-time prompts score 1 |
| D2 Invitational form | Standard or urgency-oriented form | Some calm or aesthetic qualities that are not the core of the design | Calm, slow, material or tangible form is the deliberate core of the design | Visual polish alone scores 1; a chat or LLM interface scores 0 unless its form is changed for reflection (A2) |
| D3 Interpretive openness | Verdicts, scores, labels, recommendations | Material the person interprets, framed by system goals | Designed ambiguity, juxtaposition or questions | A progress-to-goal bar scores 0; code the dominant output |

| Dimension | 0 Absent | 1 Partial | 2 Strong | Key decision rules (v1.1) |
|---|---|---|---|---|
| D4 Reflective depth | No user reflection targeted | Supports recall or description | Scaffolds questioning, explaining, alternatives or reframing | Feedback displays score 1 only if the user's reflection is a stated aim (A3); 2 needs a mechanism that asks why or what else |
| D5 Temporal span | Momentary or single-session use | Some history or trend view (days–weeks) | Designed for revisiting or anticipation over months or years | Study length does not count; a stored log never surfaced scores 0 |
| D6 Protected idleness | Fills idle moments or demands attention | On demand or peripheral; neither fills nor shapes idle time | Deliberately preserves or shapes unfilled time | 0 only for notifications, prompts, nudges or feeds; 'does not protect' is not 0 (A1) |
| D7 Ethical non-extraction | Covert manipulation, surveillance or exposure | Data practice described without a design commitment (NR if silent; scored as 1) | Explicit, embodied commitment that reflection serves the person | Research-ethics consent alone does not count as reporting |

## 5 RELIABILITY AND PROFILES OF SYSTEMS AND PARADIGMS (PHASE 2)

This section reports the results of the second phase, in which every system was scored on the seven dimensions. It first reports how reliably the dimensions could be coded, followed by how the corpus is distributed, how close each system and paradigm comes to reflective computing, how the dimensions relate to one another, and how pace, form, domain and period are associated with the scores. It closes with three checks, namely what the papers leave unreported, what the dimensions miss, and whether the findings hold among systems that aimed at reflection. Throughout, a system's proximity score (0–14) is the sum of its seven-dimension scores, and a paradigm's score on a dimension is the mean across its systems, where 0 means absent, 1 partial and 2 strong.

### 5.1 Reliability of the dimensions

Before any result can be trusted, it must be shown that the dimensions can be scored consistently by different people. Table 5 therefore reports agreement between the two independent coders before adjudication, using the measures described in Section 3.5. Invitational form ($\alpha = 0.84$) was coded reliably. Interpretive openness ($\alpha = 0.75$), tempo agency ($\alpha = 0.67$), reflective depth ($\alpha = 0.69$) and temporal span ($\alpha = 0.67$) reached the threshold for tentative conclusions, with high chance-corrected agreement on AC2 (0.75–0.90). Ethical non-extraction, however, shows the expected prevalence paradox. Exact agreement was 92% and AC2 0.97, but $\kappa$ (0.56) and $\alpha$ (0.57) are depressed because almost every system sits at one level [44]. In other words, when nearly all systems receive the same score, these two measures treat most agreement as chance, even though the coders rarely disagreed. The proximity score agreed well (ICC (2,1) = 0.82), and the second coder's scores were on average 0.4 points higher.

Table 5: Agreement between the two independent coders before adjudication (N = 134 systems).

| Code | Exact agreement | Weighted κ | Krippendorff's α (ordinal) | Gwet's AC2 | Mean, author / second coder |
|---|---|---|---|---|---|
| D1 Tempo agency | 66% | 0.66 | 0.67 | 0.75 | 0.85 / 0.88 |
| D2 Invitational form | 75% | 0.81 | 0.84 | 0.83 | 0.84 / 0.69 |
| D3 Interpretive openness | 70% | 0.76 | 0.75 | 0.78 | 0.87 / 0.86 |
| D4 Reflective depth | 78% | 0.70 | 0.69 | 0.87 | 0.65 / 0.75 |

| Code | Exact agreement | Weighted κ | Krippendorff's α (ordinal) | Gwet's AC2 | Mean, author / second coder |
|---|---|---|---|---|---|
| D5 Temporal span | 81% | 0.75 | 0.67 | 0.90 | 0.40 / 0.49 |
| D6 Protected idleness | 46% | 0.39 | 0.32 | 0.66 | 0.62 / 0.97 |
| D7 Ethical non-extraction | 92% | 0.56 | 0.57 | 0.97 | 0.93 / 0.90 |
| Pace (System/Shared/User) | 62% | 0.43 (unweighted) | – | – | – |
| Proximity score (0–14) | – | – | – | – | ICC (2,1) = 0.82 |

Two codes did not reach acceptable agreement, namely protected idleness (α = 0.32) and pace (61.9% agreement, κ = 0.43). The disagreements, however, were not random. On D6, 44 of the 72 disagreements were systems the author had scored 0 and the second coder 1, that is, on-demand apps and peripheral displays that neither fill nor protect idle time. The published rubric defines 0 as "fills or demands attention in idle moments," which supports the second coder, whereas the author's Phase 1 coding had treated the absence of protection as 0. On pace, most disagreements concerned the boundary of "shared." Two further systematic patterns appeared on D2, for calm and ambient objects whose authors argued for peripheral awareness rather than lingering, and on D4, for feedback displays that show behavior back to people without aiming at their reflection. These are decisions about the constructs, not about individual systems. Therefore, they were resolved as codebook amendments before any cell was adjudicated (Table 4, v1.1). Under these amendments, D6 follows the published rubric, and calm or material form that is the deliberate core of a design scores 2 on D2. In addition, feedback displays score 1 on D4 only when the user's reflection is a stated aim, and pace is "shared" only when both the system and the person regularly initiate interaction.

Agreement was not higher where the second coder had the full text in the first round. Exact agreement on the 85 systems coded from full text was no higher than on the remainder on any dimension, and on D6 it was markedly lower (36% versus 63%). Because the two groups contain different papers, this comparison is only indicative. Nevertheless, it suggests that disagreements reflected ambiguity in the codebook more than missing information, which is what the amendments addressed. When the 49 systems first coded from weaker sources were re-coded from the full text under v1.1, agreement with the author's original scores was higher than first-round agreement on the full corpus (Table 5) for protected idleness (α = 0.67), invitational form (0.88), interpretive openness (0.79) and reflective depth (0.73), and lower for tempo agency (0.63) and temporal span (0.60), which remain the least reliable of the six content dimensions. It is important to note that these values are not an independent replication, because the re-coding followed adjudication and used the amended codebook, so a rise in agreement cannot be attributed to the fuller source alone. The re-check did, however, change the record. The full-text code differed from the provisional consensus on 73 of the 392 dimension and pace codes for these systems (19%), and the author accepted 72 of them (the exception was a proposed rise from 1 to 2 on non-extraction). A further 14 D7 codes moved between "not reported" and 1, which leaves the score unchanged. The final scores therefore rest on the full text for all 134 systems.

### 5.2 Composition of the corpus

Before comparing systems, it is necessary to describe what the corpus contains, since every later comparison depends on how many systems each paradigm, period and form contributes. The 134 systems span all six five-year periods, with 6 from 1996–2000, 12 from 2001–2005, 19 from 2006–2010, 35 from 2011–2015, 50 from 2016–2020 and 12 from 2021–2026. The mix of paradigms changes markedly over time (Figure 2). Figure 2 is a bubble matrix (also called a balloon plot), in which each row is a paradigm, each column a period, and dot area shows how many systems fall in that cell. We use it rather than a stacked area chart or streamgraph because eleven paradigms would need eleven colors and the small paradigms (five or six systems) would shrink to slivers. In the matrix, each paradigm keeps its own row, so its rise and fall reads along the row, and empty cells stay visibly empty. Moreover, because area is judged less accurately than position [22], each bubble also carries its count. Calm and connectedness designs dominate the late 1990s, and persuasive technology rises through the 2000s. Slow, ludic and experience-centered design is present from 2004 onward, while affective computing, personal informatics, digital wellbeing and mindfulness peak in 2016–2020. After 2020, the corpus is dominated by human-AI interaction, as 11 of the 12 systems from 2021–2026 are AI systems.

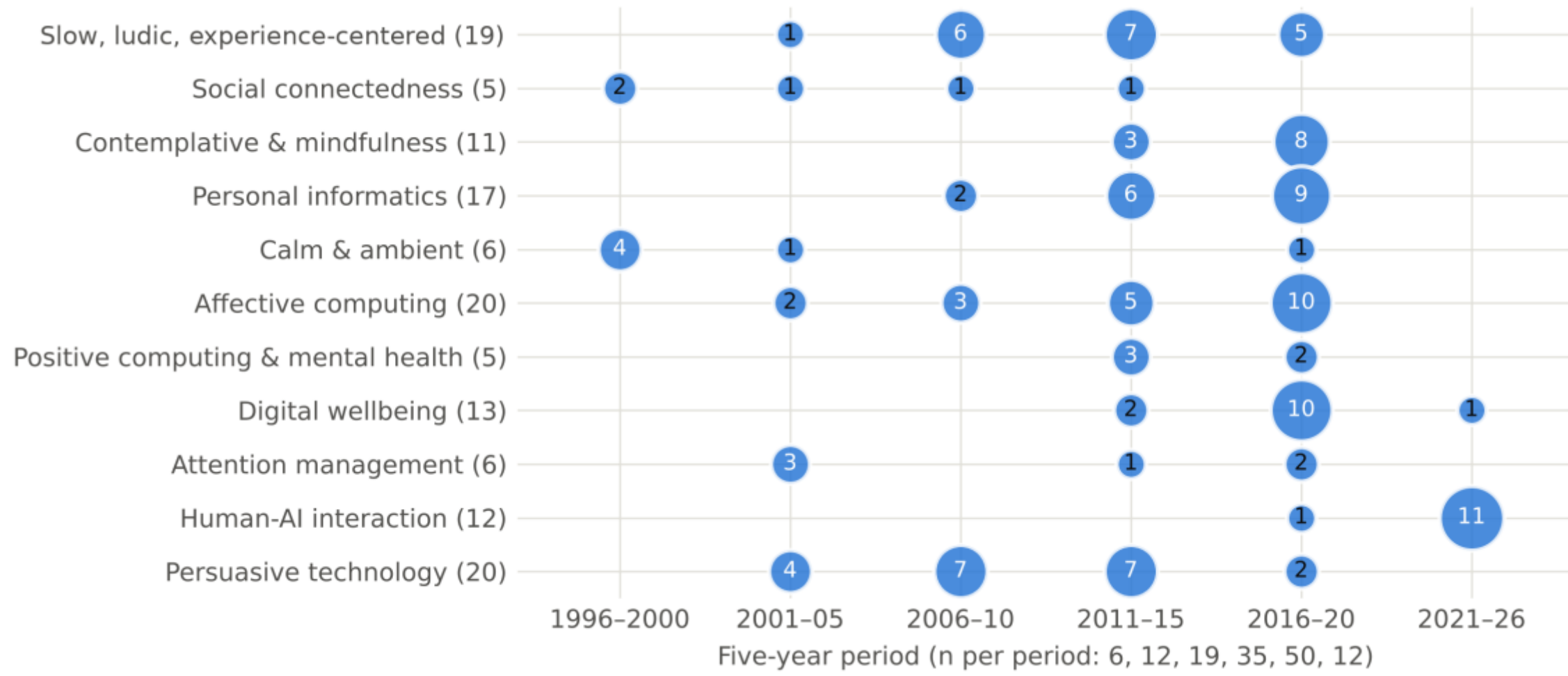

Figure 2: Systems in the corpus by originating paradigm and five-year period (N = 134), shown as a bubble matrix. Dot area and number show how many systems.

Affective computing (20) and persuasive technology (20) contribute the most systems, followed by slow, ludic and experience-centered design (19), personal informatics (17), digital wellbeing (13), human-AI interaction (12) and contemplative and mindfulness computing (11). In contrast, calm and ambient computing (6), attention management (6), social connectedness (5) and positive computing and mental health (5) are small groups. By form, 56 systems (42%) are screen-based, followed by ambient displays (21), tangible objects (17), hybrids (16), wearables (12), conversational agents (7) and immersive environments (5). The most common domains are health (44), work (26) and personal life (20).

### 5.3 How close systems come to reflective computing?

This section asks how close individual systems come to reflective computing, using the proximity score, the sum of a system's seven-dimension scores, which ranges from 0 (no reflective quality present) to 14 (every quality strongly present). The dimensions set a standard that most of these systems were not designed to meet. Hence, a corpus without a full-scoring system is expected, and what is informative is how close systems come and which dimensions they miss. Proximity scores range from 0 to 11, with a mean of 5.7 and a median of 6. No system reaches the maximum of 14, and none scores 2 on every dimension. Five systems share the highest score of 11: Personhood jewelry [176], Inner Garden [158], Olly and Slow Game [141], Chronoscope [19] and Olly field study [142]. Thirteen systems score 10 or more, and 31 score 3 or less. Only 16 systems reach at least a partial level on all seven dimensions: Affective Diary [117], Time capsule [148], Family Archive [101], Mind the Body [161], Indoor Weather Stations [59], Personhood jewelry [176], Reconsidered bathroom scale [93],

Spheres of Wellbeing [172], Counterfunctional things [151], Datacatcher [61], Inner Garden [158], Olly and Slow Game [141], Chronoscope [19], Olly field study [142], Balance, Cogito and Dott [131] and Trackly [4].

At the other end, 5 systems score 0 or 1: Playful Bottle [20], Fit4life [154], Vibration nudge [143], Voice self-perception [29] and Opinionated LLM co-writing [89]. Most of these deliberately alter or expose people's experience or behavior without their awareness or control. Fit4life [154] scores 0 by design, as it is a critical design that embodies the failings of persuasive technology to provoke reflection among designers, not users.

### 5.4 Profiles of the eleven paradigms

A paradigm profile shows, for one paradigm, its average score on each of the seven dimensions. Comparing profiles reveals which traditions already hold which reflective qualities and where each falls short. Figure 3 shows each paradigm's mean score on each dimension, and Figure 4 shows the distribution of proximity scores within each paradigm, with bootstrap confidence intervals for the mean, and Table 6 gives the values. A confidence interval gives the range within which a paradigm's true mean is likely to lie. When the intervals of two paradigms overlap substantially, the data cannot reliably tell them apart. Figure 3 is an annotated heatmap [179], that is, an 11 × 7 matrix in which cell shade encodes the mean score. A heatmap suits this question because what matters is the pattern across the whole matrix, which paradigms are strong where, rather than any single value. A grouped bar chart of 77 bars, or a radar chart of 11 overlapping polygons, would hide that pattern. The color scale is single-hue and sequential because scores run in one direction from 0 to 2, rows are ordered by mean proximity so that the ranking reads from top to bottom, and each mean is printed in its cell for exact reading. Figure 4 is a strip (jitter) plot with means and confidence intervals. Four paradigms have six or fewer systems, and with groups that small, a bar chart of means or a box plot would hide how few and how spread the scores are [178]. Plotting every system shows both, and the overlaid bootstrap 95% intervals let readers judge by eye which differences are robust [31]. The Kruskal–Wallis test (Section 3.6) shows that paradigms differ significantly (Kruskal–Wallis $H = 49.1$, $p < .001$), that is, the differences are very unlikely to be due to chance. However, the confidence intervals overlap heavily except at the extremes, so the paradigms are better analyzed as tendencies than as ranks.

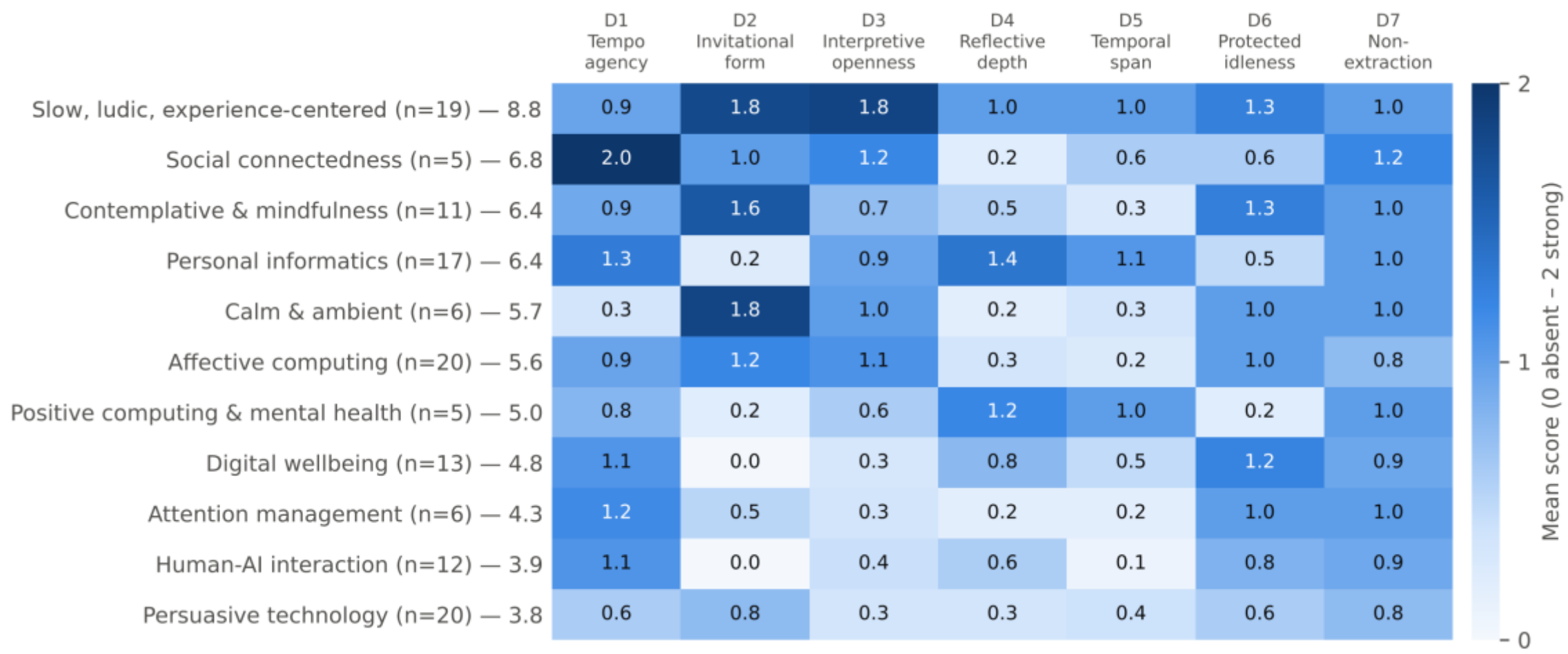

Figure 3: Annotated heatmap of mean dimension scores by originating paradigm (0 absent – 2 strong), ordered by mean proximity score, which follows each label.

At the top, slow, ludic and experience-centered design has the highest mean (8.8, 95% CI 8.1–9.6, range 6–11), and its interval overlaps only marginally with those of social connectedness (5.6–8.2) and contemplative and mindfulness computing (4.4–8.3). Its strengths are invitational form (mean 1.8 of 2), interpretive openness (1.8) and protected idleness (1.3).

A middle band follows. Personal informatics (6.4), social connectedness (6.8), contemplative and mindfulness computing (6.4), affective computing (5.6) and calm and ambient computing (5.7) each express some reflective qualities strongly and others hardly at all. For instance, connectedness systems are strong on tempo agency (2.0) but weak on depth, while mindfulness and calm systems are strong on form but have almost no temporal span. Personal informatics is their mirror image, being comparatively strong on depth (1.4) and temporal span (1.1) but almost without invitational form (0.2).

Furthest from reflective computing are digital wellbeing (4.8), positive computing and mental health (5.0), attention management (4.3), human-AI interaction (3.9, CI 2.8–5.1) and persuasive technology (3.8, CI 2.9–4.7). Persuasive technology is weak on every dimension. Human-AI interaction scores 0 on invitational form in all 12 of its systems and has almost no temporal span. However, its tempo agency (1.1) is comparatively high because writing assistants leave the person in charge of when to invoke the model.

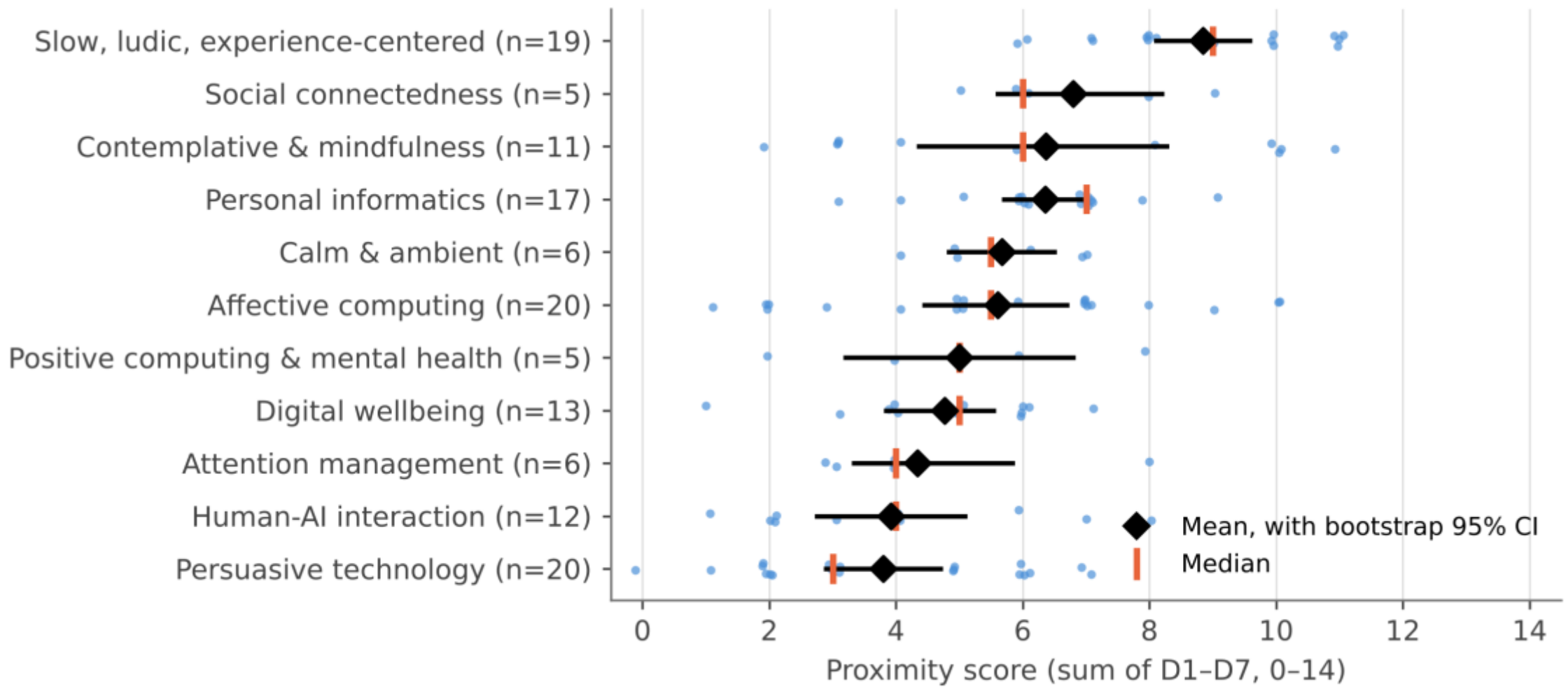

Figure 4: Proximity scores within each paradigm (strip plot with means and confidence intervals). Dots are systems; diamonds and bars show the mean and its bootstrap 95% confidence interval; ticks show the median. Four paradigms have six or fewer systems.

Table 6: Proximity scores by paradigm with percentile bootstrap 95% confidence intervals (10,000 resamples). † Six or fewer systems. Kruskal–Wallis H = 49.1, p < .001.

| **Paradigm** | **n** | **Mean** | **95% CI** | **Median** | **Range** | **System-paced** |
|---|---|---|---|---|---|---|
| Slow, ludic and experience-centered | 19 | 8.8 | 8.1–9.6 | 9 | 6–11 | 7 of 19 |
| Social connectedness † | 5 | 6.8 | 5.6–8.2 | 6 | 5–9 | 0 of 5 |
| Contemplative and mindfulness | 11 | 6.4 | 4.4–8.3 | 6 | 2–11 | 5 of 11 |
| Personal informatics | 17 | 6.4 | 5.7–6.9 | 7 | 3–9 | 2 of 17 |
| Calm and ambient † | 6 | 5.7 | 4.8–6.5 | 5.5 | 4–7 | 5 of 6 |
| Affective computing | 20 | 5.6 | 4.4–6.7 | 5.5 | 1–10 | 10 of 20 |
| Positive computing and mental health † | 5 | 5.0 | 3.2–6.8 | 5 | 2–8 | 2 of 5 |
| Digital wellbeing | 13 | 4.8 | 3.8–5.5 | 5 | 1–7 | 2 of 13 |
| Attention management † | 6 | 4.3 | 3.3–5.8 | 4 | 3–8 | 3 of 6 |
| Human-AI interaction | 12 | 3.9 | 2.8–5.1 | 4 | 1–8 | 4 of 12 |
| Persuasive technology | 20 | 3.8 | 2.9–4.7 | 3 | 0–7 | 11 of 20 |

### 5.5 Prevalence of each dimension

Whereas the previous section compared paradigms, this section compares the dimensions themselves, asking which reflective qualities are common across the whole corpus and which are rare, regardless of paradigm. The dimensions differ sharply in how often they appear (Table 7). Two are strikingly rare at the strong level. Only 12 systems scaffold reflection beyond description, and only 11 are designed for revisiting or anticipation over months or years. Moreover, temporal span is the most often absent dimension (74 systems). Protected idleness is most often *partial*, since few systems fill idle time, but few protect it either. Only two systems make ethical non-extraction an explicit design commitment.

Table 7: Distribution of consensus scores on each dimension across the 134 systems.

| Dimension | Strong (2) | Partial (1) | Absent (0) |
|---|---|---|---|
| D1 Tempo agency | 31 | 69 | 34 |
| D2 Invitational form | 43 | 29 | 62 |
| D3 Interpretive openness | 33 | 48 | 53 |
| D4 Reflective depth | 12 | 64 | 58 |
| D5 Temporal span | 11 | 49 | 74 |
| D6 Protected idleness | 18 | 84 | 32 |
| D7 Ethical non-extraction | 2 | 119 (56 not reported, 63 reported) | 13 |

### 5.6 Two cultures of reflection, through form and through depth

This section asks whether the dimensions occur independently or whether some tend to appear together. If certain qualities consistently co-occur, they point to distinct ways in which designers have supported reflection, which we call cultures of reflection. These are families of designs that share a way of inviting reflection, a set of methods and a research tradition. Two such groupings are relevant. The **form dimensions** are invitational form (D2), interpretive openness (D3) and protected idleness (D6), which concern how an artifact looks, behaves and leaves room for meaning and quiet, and their sum (0–6) is called the form cluster. The **depth dimensions** are reflective depth (D4) and temporal span (D5), which concern how far a design structures thinking and over how long a time. The analysis below tests whether these groupings are real, using correlations, factor analysis and latent class analysis (Section 3.6).

The dimensions do fall into two groups. Invitational form and interpretive openness are correlated (Spearman $\rho = 0.50$, 95% CI [0.37, 0.63]), meaning that systems strong on one tend to be strong on the other, and both are weakly associated with protected idleness (D2–D6 $\rho = 0.30$). Reflective depth and temporal span form a second group ($\rho = 0.46$, CI [0.32, 0.59]). Importantly, the two groups are independent, as invitational form and reflective depth are uncorrelated ($\rho = -0.01$, CI [-0.17, 0.16]). In other words, knowing a system's form score tells us nothing about its depth score.

Factor analysis of the polychoric correlations confirms this structure. As explained in Section 3.6, factor analysis looks for a small number of underlying factors that account for the correlations, and parallel analysis decides how many factors to keep by comparing each factor's eigenvalue, a measure of how much variation it explains, with what random data would produce. Parallel analysis retained 2 factors (eigenvalues 2.36 and 1.57 against random data 95th percentiles of 1.40 and 1.23). In the oblique two-factor solution (Table 8), invitational form, interpretive openness and protected idleness load on a *form* factor, whereas reflective depth, temporal span and tempo agency load on a *depth* factor. A loading near 1 means a dimension belongs strongly to a factor, and a loading near 0 means it does not. The factors correlate at 0.18. Interpretive openness cross-loads on both (0.62 and 0.55), suggesting that leaving meaning open is shared ground between the cultures.

Table 8: Two-factor solution of the polychoric correlations of D1–D6 (minimum-residual extraction, oblimin rotation). Loadings ≥ 0.40 in bold. Factor correlation = 0.18.

| Dimension | Form factor | Depth factor |
|---|---|---|
| D2 Invitational form | **0.97** | -0.12 |
| D3 Interpretive openness | **0.62** | **0.55** |
| D6 Protected idleness | **0.42** | 0.03 |
| D4 Reflective depth | -0.09 | **0.82** |
| D5 Temporal span | -0.08 | **0.68** |
| D1 Tempo agency | -0.03 | **0.47** |

Latent class analysis gives the same answer from a different model. Rather than grouping dimensions, it groups systems with similar score profiles into classes, and the number of classes is chosen by the BIC, which balances how well a model fits against how complex it is (Section 3.6). BIC was lowest for three classes, whose mean profiles are shown in Figure 5. A *form culture* (n = 43) scores high on invitational form and openness and contains 17 of the 19 slow-technology systems, 10 affective, 5 mindfulness and 5 calm systems. A *depth culture* (n = 43) scores higher on depth, span and tempo agency and contains 15 of the 17 personal informatics and 9 of the 13 digital wellbeing systems. A third class (n = 48) expresses neither culture and contains 12 persuasive and 9 human-AI systems. Figure 5 is a Cleveland dot plot [22] with one symbol per class on a shared 0–2 axis and a gray bar spanning the range between the classes on each dimension. We prefer it to a grouped bar chart or a radar chart because all three class means sit on one common position scale, which is the most accurately read encoding [22]. Moreover, the length of the gray bar shows at a glance where the classes separate most (invitational form and interpretive openness) and least (protected idleness and tempo agency).

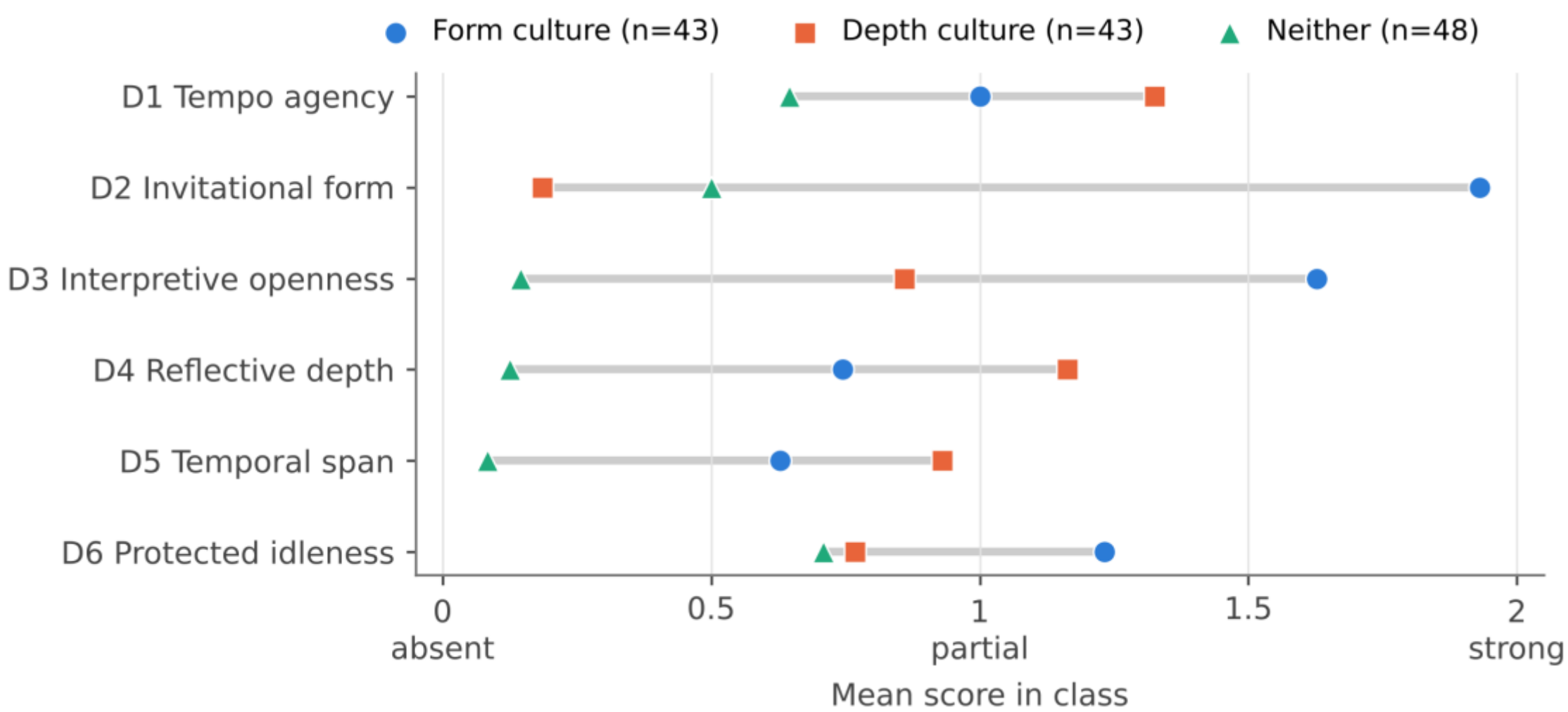

Figure 5: Mean dimension scores in the three latent classes (D1–D6), shown as a Cleveland dot plot. Each row is a dimension; symbols mark each class's mean and the gray bar spans the range between classes.

Figure 6 shows what this means in practice. It is a bubble chart of counts (a balloon plot) for two discrete scores. Because the 134 systems occupy only 19 of the 21 possible combinations of form cluster and depth, a scatter plot would stack many systems on the same point. Hence, bubble area and a printed number show how many systems share each cell, and the shaded region marks the combination that is almost empty. Of the 12 systems that scaffold reflective depth, eleven score 3 or less of 6 on the form cluster (D2 + D3 + D6), and these are screen-based apps and conversational agents such as MAHI [125], Reflection Companion [107], Robota [106] and MindfulDiary [100]. Conversely, of the 44 systems scoring 4 or more on the form cluster, only one scaffolds depth, namely Somaesthetic appreciation design [79], whose crafted artifacts are paired with voice-guided body scans that ask people to attend to and compare bodily sensations. Its depth score was one of the closest calls in the coding, so it marks where a bridge could start rather than a finished one. Thus, it can be said that HCI has developed two largely separate ways of supporting reflection. One works through the material and aesthetic qualities of things, that is, slow, ambiguous and quiet artifacts that leave room to wonder. The other works through structured thought, that is, prompts, questions, dialogues and data views that move people from describing to questioning. These two ways have almost never been combined.

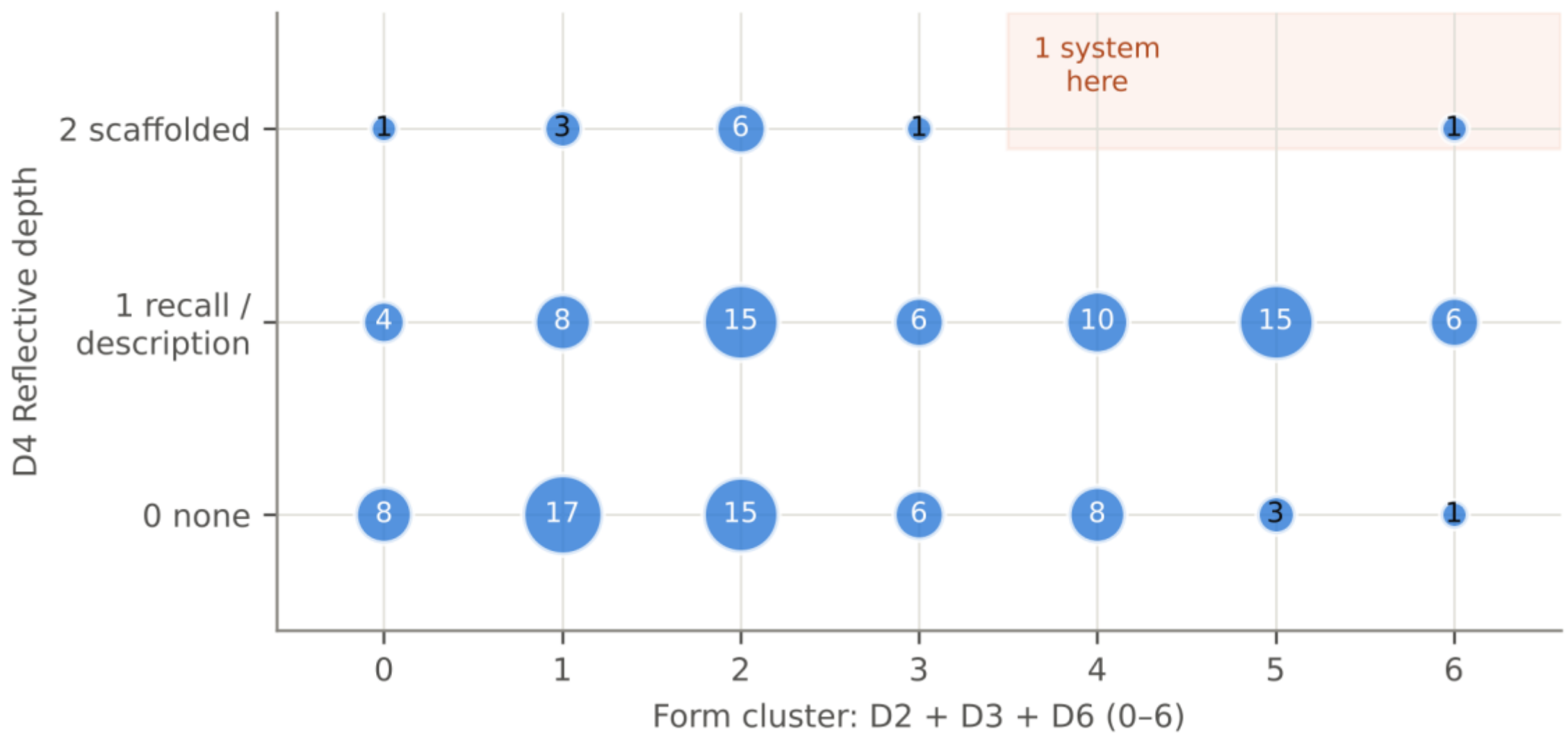

Figure 6: Systems by form cluster (D2 + D3 + D6) and reflective depth (D4), shown as a bubble chart of counts. Bubble size and number show how many systems. Only one system combines a strong form cluster (4 or more) with scaffolded depth.

Figure 7 shows the same split system by system. Each column of the UpSet plot [115] is one exact combination of dimensions on which a system scores 2 (strong). We use an UpSet plot rather than a Venn or Euler diagram because seven dimensions allow 127 possible combinations, far beyond the two or three sets a Venn diagram can show legibly. The UpSet matrix lists only the combinations that occur, sorted by frequency, so both the common and the missing combinations can be analyzed directly [115]. No system is strong on more than four of the seven dimensions, and 52 are strong on none. Moreover, apart from tempo agency alone (13), which belongs to neither culture, the most frequent combinations lie within one culture, namely invitational form alone (13), form with interpretive openness (10) and reflective depth alone (6). Twelve systems are strong on at least one dimension of each culture. However, in eight of them the depth dimension is temporal span, in slow and memory-oriented designs such as Photobox [138] and Chronoscope [19], and in three, scaffolded depth is paired only with interpretive openness, in conversational and explanation systems such as Reflection Companion [107] and Robota [106]. Only Somaesthetic appreciation design [79] combines strong invitational form with scaffolded depth.

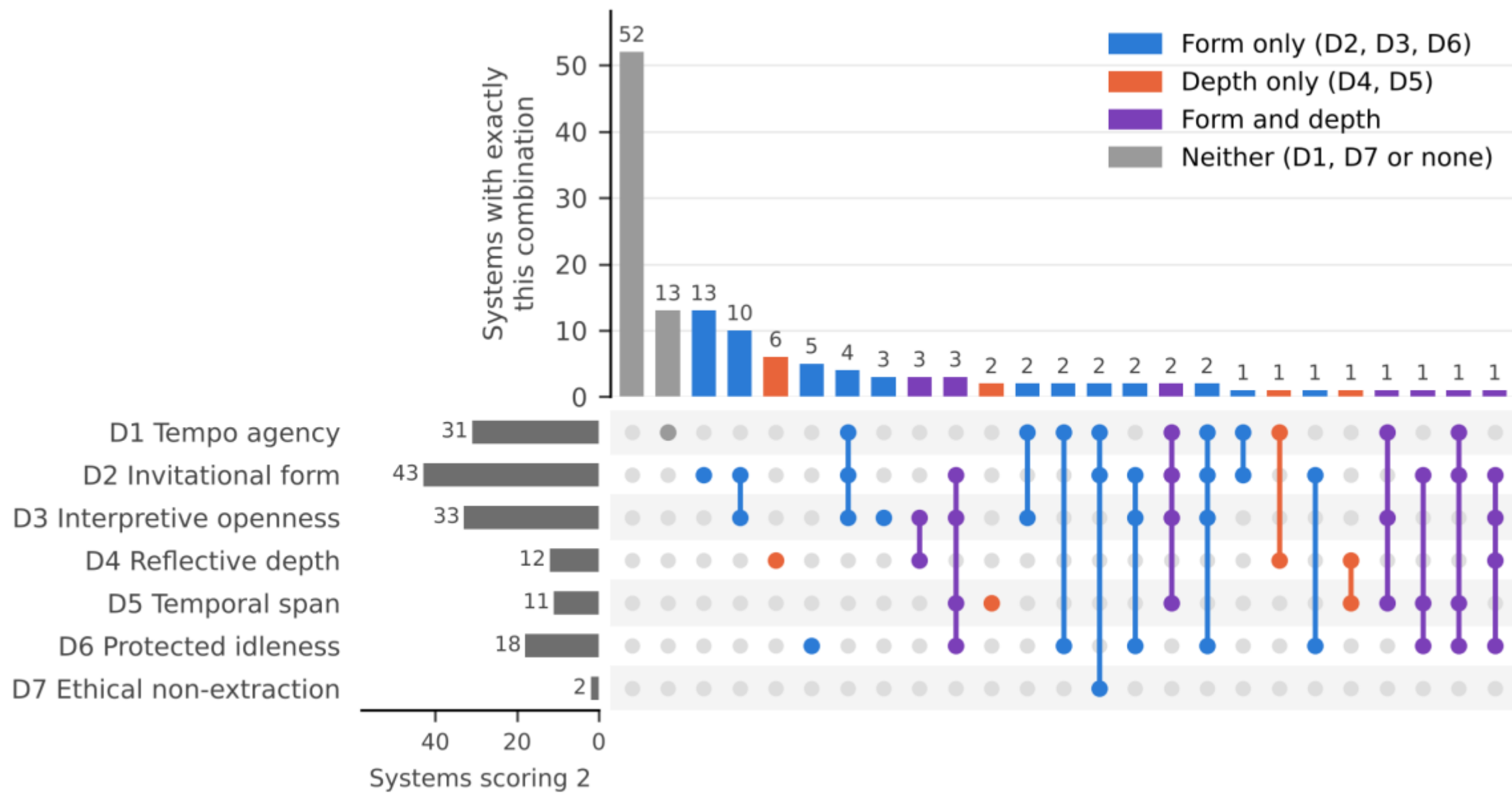

Figure 7: UpSet plot of the combinations of dimensions on which systems score 2 (strong). Bars at the top show how many systems have exactly each combination; filled dots mark the dimensions in it; bars on the left show how many systems score 2 on each dimension. Color shows whether a combination draws on the form dimensions (D2, D3, D6), the depth dimensions (D4, D5), both, or neither.

### 5.7 Who sets the pace, and why it matters?

Pace describes who decides when interaction happens. A system is coded as system-paced when it initiates and times essentially all interaction, for example through scheduled prompts or automatic displays. It is coded as shared when both the system and the person regularly initiate it, and as user-paced when the person decides when to engage. Pace is examined separately because the gap analysis suggested that control over tempo may condition the other reflective qualities (Section 4.1). The system or designer sets the pace in 51 systems (38%), pace is shared in 42 (31%) and the person sets it in 41 (31%). User-paced systems have higher mean proximity scores (7.4) than shared (5.8) and system-paced ones (4.2). Pace varies sharply by paradigm. The system sets the pace in 11 of 20 persuasive and 10 of 20 affective systems, and in 7 of 19 slow designs despite their high scores elsewhere. In these designs, slowness is offered as a gift whose timing the person cannot change.

A simple comparison of means could be misleading, because pace, paradigm and form overlap (slow designs tend to be tangible, whereas persuasive systems tend to be system-paced). Therefore, the association was tested with ordinal regression on the proximity score without D1 (Table 9), which estimates the effect of pace while holding paradigm and form constant (Section 3.6). The results are expressed as odds ratios. An odds ratio of 4.9 means that a user-paced system has 4.9 times the odds of a higher score than a comparable system-paced one. User-paced systems had 4.9 times the odds of a higher score than system-paced ones with no controls, 10.9 times controlling for paradigm, and 13.5 times (95% CI 4.9–37.6) controlling for both paradigm and form. Shared pace was also associated with higher scores (OR 5.4, CI 2.2–13.4). The advantage, however, is not uniform. Within paradigms, on the proximity score without D1, it is clear in affective computing (user or shared pace 5.8 vs system 3.5, $p = 0.043$) and mindfulness (7.7 vs 2.8, $p = 0.006$), but not detectable in slow design (7.9 vs 7.9) or persuasive technology (3.8 vs 2.7, $p = 0.203$). Pace and the dimension scores were coded by the same coders from the same papers, so some halo effect is possible, that is, a coder's overall impression of a system could color several of its scores at once. Nevertheless, the association held when each coder's own pace codes and dimension scores were analyzed separately before adjudication (user versus system pace, OR 11.8 for the author and 19.6 for the second coder, both $p < .001$). It must be acknowledged that part of the association is built into the rubric. Protected idleness scores 0 for notifications, prompts and nudges, which are usually initiated by the system, and 15 of 51 system-paced systems score 0 on D6 against 3 of 41 user-paced ones. With D6 also removed from the outcome, the association is smaller but remains large (user versus system pace, OR 8.0, 95% CI 3.0–21.6, controlling for paradigm and form, Table 9). Dimension by dimension, and with the same controls, user pace is associated with higher interpretive openness (OR 21.3, 95% CI 5.3–85.8), protected idleness (8.5, 2.4–29.8) and reflective depth (6.3, 1.8–21.9), all significant after Holm correction for testing several dimensions, but not with temporal span (3.2, 0.95–10.7) or invitational form (3.0, 0.8–11.6), whose confidence intervals include 1, the value that means no association. No user-paced system scores 0 on non-extraction (against 10 of 51 system-paced ones). Hence, that model could not be estimated, because a regression cannot compute an odds ratio when one group never takes a value. Thus, the advantage of user pace lies in openness, depth and idleness, not in form.

Table 9: Proportional-odds ordinal regression of the proximity score without D1 (0–12) and, as a robustness check, without D1 and D6 (0–10). Odds ratios with 95% confidence intervals; AIC refers to the models without D1.

| **Predictor** | **Model 1: pace** | **Model 2: + paradigm** | **Model 3: + form** |
|---|---|---|---|
| Shared vs system pace | 2.6 [1.2, 5.4] | 4.3 [1.9, 9.7] | 5.4 [2.2, 13.4] |
| User vs system pace | 4.9 [2.3, 10.4] | 10.9 [4.4, 27.0] | 13.5 [4.9, 37.6] |
| *Without D1 and D6 (0–10):* | | | |
| Shared vs system pace | 2.4 [1.2, 5.1] | 3.6 [1.6, 8.0] | 4.3 [1.8, 10.2] |
| User vs system pace | 3.9 [1.8, 8.4] | 7.3 [3.0, 18.2] | 8.0 [3.0, 21.6] |
| Paradigm controls | – | Yes (ref. persuasive) | Yes |
| Form controls | – | – | Yes (ref. screen) |
| AIC | 599.8 | 532.2 | 520.3 |

### 5.8 Differences by form, domain and period

This section examines three further characteristics of each system. Form is the physical shape the technology takes in a person's life, such as a screen-based app, a tangible object, an ambient display, a wearable, a conversational agent or an immersive environment. Domain is the area of life the system serves, such as health, work, personal life or the home. Time refers to the five-year period in which the system was published. These comparisons matter because they show whether reflective qualities depend on the medium a designer chooses, on the setting a system is built for, or on the era in which it was made, each of which points to a different explanation and a different remedy. Tangible objects score highest (mean 9.3), followed by immersive environments (7.4), conversational agents (6.4), ambient displays (5.9) and hybrids (5.4), whereas screens (4.6) and wearables (4.3) score lowest. Since 42% of the corpus is screen-based, the forms people use most are the forms that express reflective qualities least. Conversational agents are a partial exception, as they never score on invitational form, but 4 of the 7 scaffold reflective depth. By domain, systems for personal life (mean 7.6) and the home (7.4) score highest, and work systems lowest (4.2).

Mean scores by period are broadly stable from 1996 to 2020 (5.3–6.2) and lower in 2021–2026 (4.3). However, this fall cannot be interpreted as a change over time. In this corpus, period and paradigm are fully confounded. In all, 11 of the 12 systems from 2021–2026 are AI systems, and 11 of the 12 AI systems are from 2021–2026, leaving a single non-AI system to represent recent work. The AI systems do score below the corpus as a whole (a random set of 12 systems scores this low with $p = 0.006$), but they were sampled as exemplars of AI-assisted decision-making and writing, not of reflection. Thus, a systematic sample of recent AI-for-reflection systems is needed to separate the two.

### 5.9 Silence on data use

Ethical non-extraction asks whether what a person reveals while reflecting serves that person. Because many papers do not discuss data use at all, this section examines how often they are silent and whether that silence affects the results. Most papers do not say what happens to the inner states their systems collect. In all, 56 of the 134 papers (42%) are silent on the storage, access, sharing or secondary use of users' data, emotions or reflections. Moreover, 63 describe a data practice, such as local storage or user control over sharing, without making it a design commitment, and 13 involve covert influence, surveillance or exposure. To test whether the scoring rule for silence matters, the paradigm means were recomputed with silence scored as 0 instead of 1. This would lower the paradigm means but would barely change their order (Spearman $\rho$ between the two rankings = 0.97, where 1 would mean an identical order). Silence is therefore a finding in its own right, not neutrality. In a research prototype it may be harmless, but in a deployed product it leaves the question to business models (Section 7.6).

### 5.10 What the seven dimensions miss

Any set of dimensions may miss something. To check this, the second coder, who was blind to how the dimensions were derived, recorded for each system any absence relevant to reflection that the seven dimensions did not capture. Such an "other gap" note was recorded for 65 systems, and 13 of these concerned the source or the type of artifact rather than the design. Of the 52 substantive notes, 35 described an absence already captured by one of the seven dimensions or by the reflection-aim variable, most often openness (D3), non-extraction (D7) and depth (D4). The remaining 17 fell into three candidate gaps that the dimensions do not name. The first is contestable inference (7 systems), where the system infers the person's inner state and offers no way to dispute it, so the machine's reading can overwrite the person's own memory of how they felt (e.g. AffectAura [128], MoodLight [166], Mirror Ritual [156]). The second is automation that displaces noticing (5), where the system performs the noticing, logging or drafting that would have been the occasion for reflection (e.g. PlateMate [136], SleepCoacher [35], Wordcraft [182]). The third is other-directed reflection (5), where reflection is directed at another person, a place or an object rather than the self (e.g. the Digital Family Portrait [135], the Prayer Companion [57]). The first two may be facets of D3 and D4, and the third is related to the social reflection that Phase 1 did not separate as a gap. None of these reached the frequency of the seven dimensions. Nevertheless, all three are candidates for testing on new corpora (Section 7.9).

### 5.11 Robustness among systems designed for reflection

A reviewer might object that persuasive and AI systems score low simply because they never aimed at reflection. Hence, the fair test is to look only at systems that did. Reflection by the user is the primary stated aim of 22 systems and a secondary or instrumental aim of 61 more (Table 10). As expected, the 22 systems that aimed primarily at reflection score higher (mean 8.5) than those that did not aim at it (4.1). However, the gaps persist among them. Only 8 of 22 scaffold reflection beyond description, only 5 are designed for long time spans, only 3 deliberately protect idle time, 8 have no invitational form, and none makes non-extraction a design commitment. Among the 83 systems with any reflection aim, invitational form and depth remain unrelated ($\rho = -0.08$), only one system combines a strong form cluster with scaffolded depth, and user-paced systems still score higher after controlling for paradigm and form (OR 5.8, CI 1.5–22.6, and 4.6, CI 1.2–17.6, with D6 also excluded). The gaps are therefore not an artifact of including systems that never aimed at reflection. Rather, they are present in the systems built for it.

Table 10: Main results within systems whose primary stated aim is user reflection, systems where it is a secondary or instrumental aim, and systems where it is not an aim.

| | **Primary aim (n = 22)** | **Secondary aim (n = 61)** | **Not an aim (n = 51)** | **All (N = 134)** |
|---|---|---|---|---|
| Mean proximity score | 8.5 | 6.0 | 4.1 | 5.7 |
| Scaffold depth (D4 = 2) | 8 | 4 | 0 | 12 |
| Months or years (D5 = 2) | 5 | 6 | 0 | 11 |
| Protect idle time (D6 = 2) | 3 | 9 | 6 | 18 |
| No invitational form (D2 = 0) | 8 | 32 | 22 | 62 |
| No tempo agency (D1 = 0) | 1 | 14 | 19 | 34 |
| Strong form cluster and D4 = 2 | 1 | 0 | 0 | 1 |

## 6 WHAT EACH PARADIGM LACKS, AND HOW TO CLOSE THE GAP

The previous section compared paradigms by their numbers. This section turns those numbers into design guidance by reading the corpus paradigm by paradigm through the seven dimensions. For each paradigm, it describes what its systems were designed to do, what they already contribute to reflective computing, what they lack, and a translation move, that is, a concrete design change that would carry the paradigm's own strengths toward reflective computing without abandoning its original aim. Table 11 summarizes the analysis. For the four paradigms with six or fewer systems, the profile describes the exemplars rather than the paradigm.

Figure 8 summarizes where the analysis that follows leads. For each dimension, it shows the paradigm whose systems come closest to the strong level. Social connectedness leads on tempo agency (2.0) and non-extraction (1.2), calm and ambient computing on invitational form (1.8), slow, ludic and experience-centered design on interpretive openness (1.8), personal informatics on reflective depth (1.4) and temporal span (1.1), and contemplative and mindfulness computing on protected idleness (1.3). The strengths that reflective computing needs therefore already exist. However, they are spread across five paradigms, and no paradigm averages above 1.5 on more than two dimensions. On pace, form and openness, the closest paradigm is within 0.2 of the strong level, whereas on depth, span, idleness and non-extraction even the closest paradigm leaves a gap of 0.6 to 0.9. These are thus the qualities that reflective computing has to supply rather than borrow. Two of the leading paradigms, social connectedness and calm and ambient computing, have six or fewer systems, so their positions describe exemplars rather than the paradigm as a whole. Figure 8 is a dumbbell chart with a target marker, a variant of the bullet chart [45]. For each dimension, a gray line joins the corpus mean to the closest paradigm's mean, and a shaded bar runs from there to the strong level that reflective computing aims at, so the length of the bar is the gap that remains. We prefer it to a radar chart, which is often used for such profiles, for three reasons. First, the leading paradigm differs from one dimension to the next, so joining the points into a polygon would imply a single profile that no paradigm has. Second, a polygon's area depends on the order of the axes and grows with the square of the values, so it exaggerates differences. Third, gaps read along radial axes at different angles are hard to compare, whereas on one common horizontal scale they can be compared directly [22].

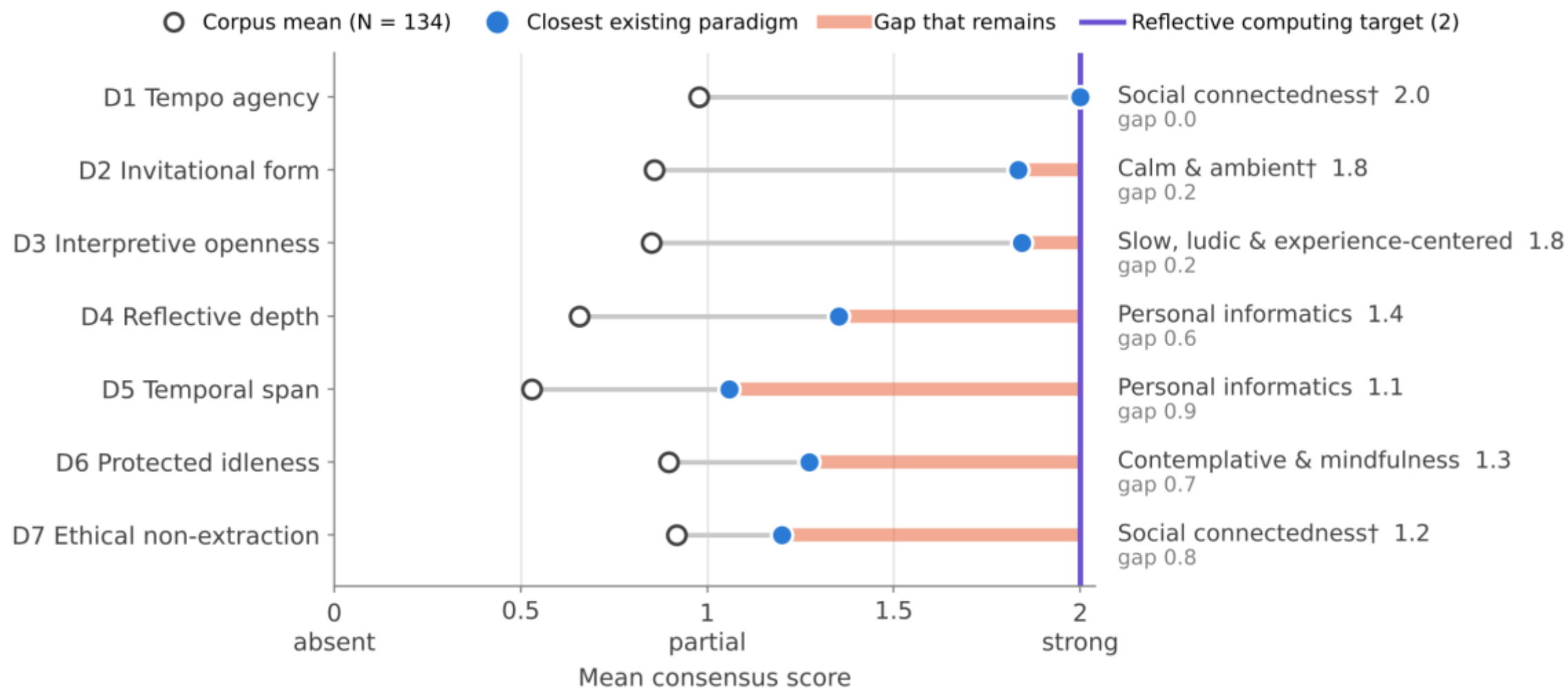

Figure 8: Where each dimension's strength already lies, and what remains open (dumbbell chart with target marker). Open circles: corpus mean (N = 134). Blue dots: the mean consensus score of the paradigm that comes closest to the strong level on each dimension, named at right. Shaded bar: the gap that remains to the strong level. Violet line: reflective computing as defined in Section 4, which aims at the strong level (2) on every dimension; it is a design target, not a measured profile. † Six or fewer systems. On D7, papers silent on data use score 1, so paradigm means cluster near 1; only 2 systems make non-extraction a design commitment.

Table 11: What each paradigm contributes, what it lacks, and the translation move toward reflective computing. † Six or fewer systems.

| Paradigm | Mean | Already contributes | Most clearly lacks | Translation move |
|---|---|---|---|---|
| Slow, ludic and experience-centered | 8.8 | Form, openness, idleness, long time spans | Tempo agency; scaffolded depth | Let people tune the slowness; add optional prompts that deepen what curiosity starts |
| Social connectedness † | 6.8 | Tempo agency; open, minimal expression | Depth; temporal span | Let traces of shared moments accumulate and be revisited together |
| Contemplative and mindfulness | 6.4 | Form; protected idleness; embodied attention | Temporal span; depth beyond the session | Connect each session to a longer arc of noticing and change |
| Personal informatics | 6.4 | Depth; temporal span; user pacing | Invitational form; protected idleness | Give data a slow, physical or ambient form and end-points that stop the feed |
| Calm and ambient † | 5.7 | Invitational, peripheral form | Tempo agency; depth | Let people bring ambient information to the center and ask what it means |
| Affective computing | 5.6 | Expressive, open form | Temporal span; non-extraction; contestability | Return sensed emotion as open, private, contestable material over time |
| Positive computing and mental health † | 5.0 | Scaffolded depth | Form; idleness | Add rest, ambiguity and pacing to structured exercises |
| Digital wellbeing | 4.8 | Tempo agency; protected idleness | Form; openness; temporal span | Turn limits into invitations to notice why one reached for the device |
| Attention management † | 4.3 | Protected focus | Depth; openness; form | Use protected time as an occasion to look back at where attention went |
| Human-AI interaction | 3.9 | Emerging prompts for thinking; invoked on demand | Form; temporal span; chosen pauses | Let people choose when the AI answers; make space to think before and after |

| Paradigm | Mean | Already contributes | Most clearly lacks | Translation move |
|---|---|---|---|---|
| Persuasive technology | 3.8 | Some ambient form | Agency; openness; depth | Replace target behaviors with questions that let people choose their own direction |

### 6.1 Slow, ludic and experience-centered design

Systems in this tradition make time and curiosity into design materials. The Drift Table offered a slowly moving aerial view steered by objects on a coffee table [53], and the History Tablecloth let traces of domestic activity linger and fade [54]. Threshold devices invited people to look out from home [56], and the Prayer Companion offered a slow stream of news to a contemplative community [57]. Photobox printed a few past photographs at unpredictable intervals across a year [138, 139], and Olly and Slow Game supported longer-term relations with everyday computational objects [141, 142]. Memory-oriented designs such as the time capsule [148], the Family Archive [101] and Chronoscope [19] made the past something to revisit with care. In addition, batch-deployed ludic devices such as the Home Health Horoscope, Indoor Weather Stations, Energy Babble and the Datacatcher invited communities to interpret ambiguous data [55, 59, 60, 61], and the Photostroller brought slowly changing images into a care home [58]. Balance, Cogito and Dott explored media for everyday reflection [131], counterfunctional things deliberately limited what devices could do [151], and personhood jewelry for people living with dementia carried meaning through crafted form [176]. Taken together, these systems contribute most of what reflective computing needs at the level of form. They score near the maximum on invitational form (1.8) and interpretive openness (1.8), protect idle time (1.3), and include most of the few systems designed for long time spans.

What slow design lacks is twofold. The first is tempo agency. In 7 of 19 systems the designer or system fixes the pace, so the person receives slowness rather than choosing it, and the unpredictable schedule of Photobox is central to its poetics but cannot be changed. The second is scaffolded depth, since these designs open a space for wonder but do not help people move from noticing to questioning. All 19 support recall or description (D4 = 1), and none reaches D4 = 2. Hence, the translation move is to let people tune the slowness, for example by choosing how often an artifact speaks, and to pair the openness of the artifact with optional, quiet prompts that deepen what curiosity starts. The memory-oriented designs, which already combine long time spans with user control, show that this is possible [19, 148].

### 6.2 Personal informatics

Personal informatics systems are designed for collecting and understanding personal data, such as sleep [21, 35, 92], food [91, 136], health conditions [4, 94, 125], activity [107], work [106, 129], lifelogs [40] and wellbeing patterns [10]. Some reconsidered familiar devices, such as the bathroom scale [93], while others supported self-experimentation [91, 112] or flexible, user-defined tracking [4, 98]. The paradigm contributes reflective depth (1.4), temporal span (1.1) and user pacing, since the system sets the pace in only 2 of 17 systems. Notably, MAHI's social scaffolding [125], Reflection Companion's conversational prompts [107], self-experimentation designs [91, 112] and visual cuts of lifelogs [40] are among the few systems that move people beyond description.

However, personal informatics lacks invitational form (0.2, with 14 of 17 at 0) and protected idleness (0.5). It treats reflection as information processing on a screen, and nothing in its form signals that slowing down is welcome. Unlock journaling even uses the idle moment of unlocking a phone for data entry [183]. Moreover, automation adds a further risk, because when the system logs or explains on the person's behalf, the act of noticing that would have prompted reflection disappears [35, 136]. The translation move is therefore to give data a slow, physical or ambient form and to give feeds natural end points, so that reflection on data feels like an invitation rather than a task. The slow-technology work on personal data shows what this could look like [142].

### 6.3 Social connectedness computing

Social connectedness systems are designed for awareness and intimacy between people at a distance. inTouch connected two people through linked rollers [13], Casablanca designed communication devices for the home [73], Babble made conversational activity socially translucent [42], ShareTable connected parents and children through a shared tabletop [181], and EatWell let community members share nutrition-related memories [65]. Tempo agency is this paradigm's strongest dimension (2.0), as people choose when and how to reach out. Furthermore, minimal, open channels such as tangible rollers leave meaning to the people involved.

The paradigm lacks depth (0.2) and temporal span (0.6). Most systems support the moment of contact but not reflection on what the relationship means over time. Thus, the translation move is to let the traces of shared moments accumulate and become material that people can revisit together, as Babble's persistent conversation partly did [42].

### 6.4 Contemplative and mindfulness computing

Contemplative and mindfulness systems are designed to guide attention to breath, body and present experience. Sonic Cradle connected respiration to music in a dark immersive space [175], and Spheres of Wellbeing supported mindfulness practice for women in secure hospital care [172]. Somaesthetic appreciation design developed artifacts for attending to the body [79], Inner Garden linked inner states to a mixed-reality sandbox [158], RelaWorld adapted a virtual meditation space to brain activity [109], and other systems paced breathing through light, sound, vibration or games [62, 130, 133, 145, 146, 153]. The paradigm contributes invitational form (1.6) and protected idleness (1.3), and several of its systems are among the highest scoring in the corpus [79, 158, 175].

However, temporal span is absent in 9 of the 11 systems, and reflective depth beyond the session is rare (0.5). Several breathing systems set the pace for the person [62, 133, 145], and some entrain breathing below awareness, which leaves nothing for the person to reflect on. Therefore, the translation move is to connect each session to a longer arc of noticing, that is, what changed across weeks of practice, what the person learned about their own attention, and when they choose to return.

### 6.5 Affective computing

Affective computing systems are designed for sensing, expressing, sharing or regulating emotion. Some are expressive, such as LumiTouch [18], eMoto [168], the Affective Diary [117] and the Sensual Evaluation Instrument [86]. Some share physiology between people, such as heart rate sharing [165], MoodLight [166], Breeze [48], HeartChat [70], Animo [118] and color-changing clothing [83]. Some reflect sensed affect back to the person, such as AffectAura [128], Mind the Body [161], Mirror Ritual [156] and Ripple [84], while others sense or regulate affect directly, often without the person's own interpretation [28, 29, 30, 69, 72, 119]. The paradigm contributes expressive and open form (D2 1.2, D3 1.1), and its most reflective designs give people ambiguous material to interpret rather than labels [83, 84, 117].

Affective computing, however, lacks temporal span (0.2, with 15 of 20 at 0) and non-extraction (0.8). It is the most polarized paradigm, with scores from 1 to 10. Several systems measure emotion without returning it to the person, regulate it through deliberately false feedback [28] or covert manipulation [29], or display it without the wearer's control [83]. Moreover, even where emotion is returned, the machine's inference is rarely open to dispute [128, 156, 166]. The translation move is thus to return sensed emotion to the person as open, private and contestable material that accumulates over time, rather than as a label or a covert intervention. The Affective Diary and Ripple show parts of this move [84, 117].

### 6.6 Calm and ambient computing

Calm and ambient systems are designed to move information to the periphery, as in ambientROOM [87], the water lamp and pinwheels [32], Audio Aura [134] and the Information Percolator [71]. The Digital Family Portrait conveyed an elder's wellbeing to relatives at a glance [135], and Physikit turned environmental data into physical ambient objects [82]. The paradigm contributes an invitational, peripheral form (1.8) that does not demand attention.

However, it lacks tempo agency (0.3) and depth (0.2). Calm technology reduces demand but does not invite thought. The system decides what reaches the periphery and when, and reflection, where it occurs, is often about someone else [135]. Hence, the translation move is to let people bring ambient information to the center when they choose, and to accompany it with an open question about what it means. Physikit, which lets households configure what their objects show, points in this direction [82].

### 6.7 Digital wellbeing and self-control

Digital wellbeing and self-control systems are designed to help people regulate their device use, through personal goals [74], group limits [97, 104, 105], parental mediation [103], lockouts [99], vibration nudges [143], framing [96], removing notifications [150] or newsfeeds [124], adaptable commitment interfaces [121], and design for intentional media use by children [75, 76]. The paradigm contributes tempo agency (1.1) and protected idleness (1.2), for instance when a newsfeed, autoplay or notifications are removed [76, 124, 150].

However, it lacks invitational form (0.0), interpretive openness (0.3) and temporal span (0.5). Self-control tools slow people down to stop them, and rarely to help them understand why they reached for the device or what they wanted from it, while group designs often frame non-use as a competition. The translation move is therefore to turn limits into invitations to notice. At the moment of friction, the system can ask a quiet, open question rather than enforce a rule. Plan & Play, which asks children and parents to plan media use together in advance, comes closest [75].

### 6.8 Positive computing and mental health

Positive computing and mental health systems are designed to support wellbeing and mental health, as in Echo's reflection on past positive experiences [85], EmotiCal's emotion tracking for behavior change [77], Pocket Skills for dialectical behavior therapy [163], adaptive stress reduction [108] and peer support chats [137]. The paradigm contributes scaffolded depth (1.2). For instance, Echo asks people to explain why past experiences were positive, and peer support chats ask them to question their experiences together [85, 137].

However, it lacks invitational form (0.2) and protected idleness (0.2), because these systems deliver structured exercises on screens, often with reminders, in the same form as the applications they compete with for attention. Hence, the translation move is to add rest, pacing and ambiguity to structured exercises, so that the form of a wellbeing system differs from the form of an engagement-driven app.

### 6.9 Attention and interruption management

Attention and interruption management systems are designed to protect focus and manage notifications, as in Sideshow [15], Scope [173], Lilsys [9], an email cut-off intervention [126], FlowLight [186] and a conversational agent supporting detachment from and reattachment to work [180]. Their contribution lies in some protection of focus and of the boundaries of the working day.

They lack depth (0.2), openness (0.3) and form (0.5), since protected time is treated as a resource for productivity rather than for reflection. The detachment agent is the exception, as it asks workers to reflect at the boundary of the working day [180]. Thus, the translation move is to use protected time as an occasion to look back at where attention went and why, not only as a shield against interruption.

### 6.10 Persuasive technology

Persuasive systems are designed to change behavior toward a target, such as physical activity [17, 25, 26, 122, 160], water and energy conservation [3, 43, 50, 66, 111], waste [171], transport [49], water intake [20], group participation [37, 95], movement [88, 157] and health recommendations [6, 155]. Fit4life embodies the paradigm's failings as critical design [154]. The paradigm contributes some ambient form, such as the sculpture of Breakaway [88], the Power-Aware Cord [66] and twinkly lights that lure people to the stairs [157].

However, it lacks tempo agency (0.6), openness (0.3) and depth (0.3). The system sets the pace in 11 of 20 systems, success is defined by a target behavior chosen by the designer, and social persuasion can compromise privacy, as when photographs of household waste were shared on social media [171] or drinking records were made visible to co-workers [20]. The translation move is therefore to replace target behaviors with questions that let people choose their own direction. The framing of BinCam as "reflection and social persuasion" [171] shows that the two can be combined. Nevertheless, the reflective half needs the person's own pace and privacy.

### 6.11 Human-AI interaction

Human-AI interaction systems are designed to support people deciding or writing with AI, through explanations and confidence displays [5, 174, 184], cognitive forcing functions [14], AI assistance without direct recommendations [51], question-framed explanations [34], large language model (LLM) writing tools [33, 113, 182], opinionated language models [89], procrastination intervention [2] and LLM-supported journaling [100]. The paradigm contributes emerging prompts for thinking. Cognitive forcing [14], question framing [34] and continuous summaries that help writers see their own text [33] all slow the step from AI output to human acceptance, and MindfulDiary scaffolds reflective journaling through dialogue [100]. Moreover, writing assistants that respond only when invoked give people comparatively high tempo agency (1.1).

However, all 12 systems score 0 on invitational form, and only one has any temporal span. Where AI systems slow people down, as cognitive forcing does, the pause is imposed rather than chosen, and where they fill the blank page, the act of drafting that would have prompted reflection is done for the person [182]. One system shows the opposite risk, as language models with built-in opinions shifted users' own views without their awareness [89]. Hence, the translation move is to let people choose when the AI answers and to make space to think before and after it does, through pacing controls for AI help, interface forms that signal a moment to think rather than a race to accept, and records that let people revisit what they accepted and why. This fits evidence that confidence in AI is associated with less critical thinking [114] and that generative AI changes the metacognitive work people must do [169].

## 7 DISCUSSION

This section steps back from the corpus to ask what the gaps mean. It first considers what kind of paradigm reflective computing is and how it relates to the paradigms from which it was derived and then discusses four issues that the analysis brings into focus, namely the split between two cultures of reflection, the place of tempo agency, generative AI, and reflection as something that can be extracted. It closes with design directions, an approach to evaluation, a research agenda, limitations and implications.

### 7.1 Reflective computing as a paradigm

Having reported what the corpus shows, we now turn to what it means. The analysis supports three claims about reflective computing as a paradigm. The first claim is that it is defined by what is missing, not by what is new. None of the seven dimensions is absent from the corpus at the partial level, and six of the seven are expressed strongly by at least 11 systems. What no paradigm does is hold all seven together as its aim. Slow technology holds form, openness and idleness, while personal informatics and positive computing hold depth. Connectedness and digital wellbeing hold tempo agency, memory-oriented designs hold temporal span, and only two systems hold non-extraction as an explicit commitment. Thus, reflective computing is best understood as a synthesis that names a whole the field has so far built only in parts.

This is a stronger basis for a paradigm than a novel technique would be. A paradigm proposed from a new technique must first show that its qualities can be built. Reflective computing does not carry that burden, since most of its dimensions already have working exemplars, published and evaluated at ACM venues over three decades. Its claim is, therefore, narrower and easier to test, namely that these qualities belong together, that a design expressing several of them is more reflective than one expressing a single quality strongly, and that the traditions which developed them separately have something to learn from each other.

The second claim is that reflective computing differs from its neighbors in its aim and its center. Each neighboring paradigm organizes design around a system function or an outcome, such as affective computing around emotion, persuasive technology around behavior change, calm technology around attention, positive computing around wellbeing, digital wellbeing around self-control, and human-AI interaction around joint performance. Reflective computing, in contrast, organizes design around a human capacity, namely the capacity to pause, to consider and to make sense of experience in one's own terms [36, 162]. Hence, its success criterion is not a measured outcome such as reduced stress, steps taken or decisions improved, but whether people had the time, the form and the material to think for themselves. Furthermore, it places the person's control of tempo at its center (Section 7.4).

The third claim is that reflective computing is a paradigm of qualities, not of devices. Reflective qualities appear in tangible objects, ambient displays, wearables, screens, conversational agents and immersive environments. No form is excluded and none guarantees reflection. Tangible systems score highest on average (9.3), yet the Power-Aware Cord, a glowing physical cord, scores only 7 because it persuades rather than invites [66], and conversational agents, which never score on invitational form, include 4 of the 12 systems that scaffold reflective depth. Reflective computing is not a category of hardware, as ubiquitous or tangible computing partly were, but a set of qualities that any form can carry to a greater or lesser degree. This is why it can serve as a lens on existing paradigms rather than only as a label for new systems.

### 7.2 Relation to neighboring paradigms

Since the dimensions were derived from the gaps in other paradigms, reflective computing has a specific relationship with each of them. It is not an alternative to these paradigms and does not ask designers to abandon their aims. Rather, it identifies what each paradigm already contributes to reflection and what it would need to add (Table 12).

Table 12: How reflective computing relates to the paradigms from which its dimensions were derived.

| Paradigm | Shared ground with reflective computing | Where reflective computing diverges |
|---|---|---|
| Slow, ludic and experience-centered design | Form, openness, idleness and longtime spans as design materials | Slowness should be chosen, not given; wonder should be able to deepen into reflection |
| Calm and ambient computing | Peripheral form that does not demand attention | Reducing demand is not the same as inviting thought; people should be able to bring the periphery to the center |
| Contemplative and mindfulness computing | Pausing; attention to the body and present experience | Reflection also looks back and ahead; sessions should connect to longer arcs of change |
| Social connectedness computing | People choose when and how to reach out | Shared moments become material for reflecting on relationships over time |
| Affective computing | Emotion as material for self-understanding | Emotion is returned to the person, privately, as open material; it is not measured or steered covertly |
| Personal informatics | Reflection on personal data over time; user-paced use | Data reflection should have a form that invites pausing and should not fill idle moments |
| Positive computing and mental health | Scaffolded depth toward flourishing | Wellbeing tools should differ in form and pace from the engagement-driven apps they compete with |
| Digital wellbeing and self-control | Agency over technology use | Friction is an occasion for understanding, not only for stopping |
| Attention management | Protecting time from interruption | Protected time can be used for reflection, not only for productivity |
| Persuasive technology | Concern with everyday change | Direction is chosen by the person; questions replace targets |
| Human-AI interaction | Concern with how people reason with machines | People decide when the AI answers; space is protected before and after it does |

Two patterns stand out. First, reflective computing is closest to slow technology, from which it borrows the most, but it corrects slow technology where the corpus shows it to be weakest, namely control of pace. Second, its relationship with persuasive technology and human-AI interaction is the most demanding. In both, the system or designer typically decides what counts as success, and reflective computing asks that this decision be returned to the person. This is not a rejection of behavior change or AI assistance. A persuasive system can invite people to choose their own direction, and an AI system can support reflection, as question-framed explanations [34] and MindfulDiary [100] show. It does, however, require a change in who sets the goal.

The paradigms in the corpus are not the only relatives of reflective computing. Three bodies of work also take reflection as their subject, and an informed reader will rightly ask what reflective computing adds to them. These are reflective design [164], slow technology [68] and reflective informatics [8]. Reflective computing draws on all three, and several of its dimensions are built directly on their insights. It differs from each of them, however, in what the reflection is about, in who controls the slowing down, and in how far the resulting qualities can be assessed.

Reflective design, as proposed by Sengers and colleagues, uses critical reflection to bring to the surface the unconscious values and assumptions built into technology, for designers and users alike [164]. Its principles, such as using ambiguity to invite interpretation, inform interpretive openness (D3) in this article. The difference lies in the object of reflection. In reflective design, people chiefly reflect on technology and on the culture around it, whereas in reflective computing they reflect on their own experience, beliefs and actions, with the technology serving as the means rather than the subject. Moreover, reflective design is a stance and a set of principles for the design process, while reflective computing describes the qualities of the resulting system and makes each of them assessable through a codebook.

Slow technology, proposed by Hallnäs and Redström, is the closest relative. It asks designers to make technology that takes time to understand and that creates moments of mental rest and reflection [68], and it informs invitational form (D2) and protected idleness (D6). Reflective computing departs from it in two respects that the corpus makes visible. First, slow technology usually decides the pace on the person's behalf, and in 7 of the 19 slow designs the system or designer sets it (Section 7.4). Reflective computing treats slowness as valuable only when people choose it, and it therefore places tempo agency (D1) at its center. Second, slow technology opens a space for wonder but rarely helps people move from noticing to questioning, as none of its 19 systems scaffolds reflection beyond description (Section 6.1). Reflective computing asks for both, and it adds ethical non-extraction (D7), which slow technology does not address.

Reflective informatics, developed by Baumer, identifies breakdown, inquiry and transformation as conceptual dimensions of reflection and uses them to guide the design of technologies for reflection [8]. Together with the levels of reflection described by Fleck and Fitzpatrick [46], it informs reflective depth (D4). Its dimensions, however, describe the process of reflection itself, that is, what happens in a person's thinking. Reflective computing is concerned with the conditions that a system creates around that process, namely who controls its pace, what the form signals, how long the system supports it, whether idle moments are protected, and whom the person's reflections serve. In this sense, reflective informatics describes what reflection is, whereas reflective computing describes what a system must do so that reflection can happen.

In short, reflective computing differs from these three bodies of work in four ways. It concerns the person's reflection on their own life rather than on technology. It makes the person's control over pace the central condition rather than a matter of design taste. It combines the form-based and the thought-based traditions that have so far developed apart, and it treats the ethical use of what people reveal as part of the design itself. Finally, it turns these qualities into seven dimensions that can be scored, compared and tested, which none of the three offers.

It is equally important to state what this proposal does not yet show. No system in the corpus expresses all seven dimensions strongly, and the one design that combines a strong form cluster with scaffolded depth, Somaesthetic appreciation design [79], does so only on a close call (Section 5.6). Reflective computing has therefore not yet been demonstrated in a system built for it. Moreover, the dimensions were derived from and tested on the same 134 systems, so they have not yet been applied to systems outside the corpus (Section 7.10). These are normal limits for a first articulation of a paradigm, and they define the next two steps. The first is to build and deploy systems designed to express all seven dimensions, beginning with designs that bridge form and depth (Section 7.3). The second is to apply the codebook independently to new systems. Until both steps are taken, reflective computing is best analyzed as a well-grounded proposal rather than an established paradigm, and its value will depend on whether others can use it to build and evaluate systems.

### 7.3 Bridging the two cultures of reflection

The clearest structural finding is that HCI has developed two separate ways of supporting reflection (Section 5.6), which we call two cultures of reflection. These are two families of designs, each with its own way of inviting reflection, its own methods and its own research tradition. The evidence for this is threefold, namely the correlations, a two-factor structure retained by parallel analysis, and three latent classes in which form and depth occupy different classes. The first culture works through **form**. It is found in slow technology, mindfulness, calm and expressive affective computing, mostly in tangible, ambient and immersive artifacts. Its systems invite people to slow down and wonder through material, motion, ambiguity and quiet, but rarely help them take the next step from noticing to questioning. Indeed, all 19 slow-technology systems support recall or description, and none scaffolds reflection beyond it. The second culture works through **scaffolded thought**, that is, through prompts, questions and structured exercises that guide a person's thinking step by step. It is found in personal informatics, positive computing, digital wellbeing and, more recently, conversational agents and AI. Its systems prompt, question and structure reflection through screens and dialogues that look and behave like any other application. Of the 12 systems that scaffold depth, only one scores 4 or more of 6 on the form cluster. Thus, the two cultures are not opposed. Rather, they have almost never met. The same separation holds among systems that aimed at reflection (Section 5.11), so it is not an artifact of mixing reflective and non-reflective designs.

There are good reasons why the two cultures developed apart. The first culture grew out of design research, with its emphasis on artifacts, aesthetics and ambiguity as a resource [52, 68]. Its methods favor field deployments of research products whose meaning is left open [140], and it has often been wary of telling people what to think. The second grew out of behavioral and health research, with its emphasis on measurable outcomes and evidence-based techniques [102, 116]. Its methods favor prompts whose effects can be measured, and it has often been less concerned with the form those prompts take. As a result, the strength of each culture is the blind spot of the other.

Reflective computing needs both cultures. Form without depth risks leaving people at the threshold of reflection, as an artifact can invite a pause without helping the pause become understanding. Depth without form risks asking for reflection in the same register as every other demand on attention, as a reflective prompt that arrives as a notification competes with the notifications around it and offers no signal that this moment is different. Interpretive openness, which loads on both factors, is the natural meeting point, since leaving meaning open is valued by both cultures. The one design in the corpus that scores on both is Somaesthetic appreciation design [79], whose crafted artifacts are paired with voice-guided body scans that ask people to attend to and compare bodily sensations. Its depth score was a close call. Hence, it suggests where a bridge could begin rather than showing a finished one. Similarly, Ripple, which pairs an ambiguous worn display with alternative interpretations of biosensed emotion [84], points in the same direction.

Other bridges can be imagined from the corpus. A slow artifact such as Photobox [138, 139] could offer, on request, a quiet question about the photograph it has just printed. A reflective journal such as MindfulDiary [100] could take a physical or ambient form that signals a different kind of time from the phone on which it runs. A personal informatics system could present its data through an object that changes slowly, as slow-technology work on personal data has begun to do [142]. These are design hypotheses, not findings. Nevertheless, the gap between the two cultures is where reflective computing is most likely to produce new kinds of system, and it is crucial to articulate strong concepts that bridge them [78].

### 7.4 Tempo agency as the central condition

Of the seven dimensions, tempo agency distinguishes reflective computing most clearly from the tradition closest to it. Slow technology has long argued that technology should take time and give time [68]. In practice, however, slow technology often takes time on the person's behalf. In 7 of 19 slow-technology systems, the designer or system sets the pace. Photobox prints at intervals the household cannot change [138, 139], and Olly turns at a speed fixed by its makers [141, 142]. Slowness here is a gift whose timing the recipient cannot alter.

There are good reasons for this. Unpredictability and fixed pace are part of the poetics of these designs, and a slow artifact whose pace could be adjusted at will might lose the quality that makes it slow. Indeed, within slow design, user-paced and system-paced systems score alike (Section 5.7). Across the corpus, however, imposed pace has costs. User-paced systems score markedly higher than system-paced ones on the other dimensions taken together, most clearly on interpretive openness, reflective depth and protected idleness, though not on invitational form. The advantage survives controls for paradigm and form (OR 13.5, and 8.0 when protected idleness, which depends partly on pace by definition, is also excluded), appears in each coder's independent codes, and holds among systems that aimed at reflection. Moreover, research on digital self-control has found that imposed friction is often overridden or abandoned [123, 132], and research on the sense of agency has found that features people initiate support agency better than features a system pushes [120]. Thus, a pause that is imposed may produce compliance or irritation rather than reflection.

Reflective computing therefore treats tempo agency not as one quality among seven but as the condition under which the others become reflective. Invitational form invites a pause, and tempo agency lets the person accept or decline. Protected idleness preserves unfilled time, and tempo agency lets the person decide whether to fill it. Reflective depth scaffolds questioning, and tempo agency lets the person decide when to go deeper and when to stop. The factor analysis does not single it out in this way, since tempo agency loads moderately and only on the depth factor (0.47). Its role as a condition therefore rests on the regression results rather than on the factor structure. It is important to note that reflective computing does not claim that slower is always better. A person who chooses to move quickly has exercised tempo agency, and a reflective technology should let them. What it resists is the removal of that choice, whether by an engagement-driven interface that never lets people stop or by a well-meaning one that never lets them go.

The design challenge, therefore, is to give people control of tempo without losing what makes slow designs work. The corpus offers three approaches. The first is setting a rhythm in advance and letting the system keep it, as personal goals and limits do [4, 74]. The second is choosing between quick and considered modes in the moment, as SwitchTube does [121]. The third is having the system follow a rhythm the person already has, as Sonic Cradle follows breath [175] and the memory-oriented designs let people decide when to open, revisit or add [19, 148].

### 7.5 Reflective computing and generative AI

A tempting reading of the period trend is that the newest generation of systems, built around AI, is the furthest from reflective computing. However, the consensus coding and the structure of the corpus do not support that reading in its strong form. Human-AI systems do score low (mean 3.9, CI 2.8–5.1), but not lower than persuasive technology, and the corpus cannot separate a change over time from a change of paradigm, since almost all recent systems are AI systems and almost all AI systems are recent (Section 5.8). Moreover, the twelve AI systems were sampled as exemplars of AI-assisted decision-making and writing rather than of reflection. What the corpus does show is narrower and more robust. Every AI system scores 0 on invitational form, only one supports reflection over time, and where AI systems slow people down, they usually impose the pause.

This is not because AI researchers are unconcerned with reflection. Several systems aim to slow the acceptance of AI output and keep people thinking, such as cognitive forcing functions [14], AI assistance without direct recommendations [51], question-framed explanations [34], continuous summaries that help writers see their own text [33], and LLM-supported journaling [100]. Two of these scaffolds reflective depth. The problem, however, is that they pursue reflection through only one or two dimensions. They add depth in a form visually and temporally indistinguishable from the rest of the interface, they slow people down by imposing a pause, and they are designed for the task in front of the person rather than for reflection across time.

The stakes are high. Survey evidence suggests that knowledge workers who are more confident in generative AI report less critical thinking [114], and generative AI changes the metacognitive work people must do, that is, deciding what to ask, how to judge an answer and when to rely on it [169]. Moreover, the corpus contains a direct demonstration of the risk to non-extraction, as language models with built-in opinions shifted users' views during co-writing without their awareness [89]. Guidelines for human-AI interaction have focused on making AI behavior understandable and correctable [1], which is necessary but does not address whether people have the time and space to think.

The seven dimensions suggest what reflective AI might look like. Under tempo agency, people would choose when the AI answers, that is, before they have formed their own view, after it, or not at all. Under invitational form, the interface would signal when a moment calls for thought rather than presenting every response with the same urgency. Under interpretive openness, the AI would offer alternatives, questions and juxtapositions as well as answers, as question-framed explanations do [34]. Under reflective depth, it would help people examine why they accept or reject a suggestion. Under temporal span, it would keep a record that lets people revisit what they accepted and how their thinking changed. Under protected idleness, it would not fill every pause with a suggestion. Under ethical non-extraction, it would not use people's hesitations and reflections to steer their opinions or to train systems without their knowledge. Furthermore, the open coding adds a caution. When the model drafts, logs or explains on the person's behalf, it can remove the very occasion for reflection (Section 5.10).

### 7.6 Reflection as something that can be extracted

Ethical non-extraction is the least addressed dimension. Only two systems make it an explicit, embodied design commitment, 13 systems score 0, and 56 papers are silent on what happens to the inner states they collect. This might suggest that non-extraction is a minor concern. However, the analysis suggests otherwise, for three reasons.

First, the systems that score 0 are not marginal. They include affect and engagement sensing that does not return what it measures to the person [69, 72], interventions that alter emotional experience through false physiological feedback [28, 62] or covert voice manipulation [29], public display of individuals' behavior in groups and workplaces [20, 37, 97], social persuasion that exposed household waste on social media [171], and opinionated language models [89]. Several of these are technically sophisticated and widely cited. In other words, the technologies needed to sense, influence and expose inner states are mature.

Second, the moments reflective computing seeks to protect are exactly the moments most valuable to extract. A pause before a decision reveals hesitation, a journal entry reveals feelings and intentions, and an idle moment is an opportunity to capture attention. Interfaces designed to exploit fast, automatic choice are well documented [64], and as sensing and AI become more capable, the same techniques can be applied to slower and more private moments. As a result, a technology that invites reflection without protecting it may create new material for extraction.

Third, silence is not a stable position. In all, 42% of the papers in the corpus do not say what happens to the reflections, emotions or pauses their systems collect. In a research prototype this may be harmless, but in a deployed product the same silence leaves the question to business models. Reflective computing therefore treats non-extraction as a positive design commitment rather than the absence of harm. Reflections should be private by default, people should know what is kept and why, and nothing revealed while reflecting should be used to persuade, profile or engage them without their knowledge. In addition, the open coding adds contestability, since when a system infers a person's inner state, the person should be able to dispute it. Two systems show that such commitments can be built into the artifact rather than left to a privacy policy. The designers of Casablanca removed automatic presence sensing after focus groups described it as surveillance, so that presence is shared only by deliberate action [73], and the private recordings of the Spheres of Wellbeing can be accessed only through their owner's sphere [172]. The critical alternatives of Ripple [84], in turn, make the question of who benefits from emotional data part of the design itself.

### 7.7 Design directions for reflective computing

Six design directions follow from the gaps. The first is to combine form and depth. Reflective prompts can be given an invitational form, and slow artifacts can be given optional scaffolds for deeper thought (Section 7.3), so that the material qualities of an artifact and the questions it poses support each other. The second is to make slowness a choice. Pace controls, declinable pauses, and quick and considered modes of the same task let people choose to slow down rather than being slowed down (Section 7.4). Moreover, personal informatics and digital wellbeing already show that people use such controls [4, 74, 121].

The third direction is to leave meaning to the person. Scores, verdicts and labels can be replaced with material, juxtapositions and questions wherever the meaning of an experience is for the person to decide, and machine inferences can be made contestable. Sensed emotion, personal data and AI output are where the corpus most often closes down meaning that could be left open [34, 83, 117]. The fourth is to design for the long run, by building systems that accumulate meaning, support revisiting and anticipation, as memory-oriented designs do [19, 148], and allow graceful lapse and return. Hence, such systems should be evaluated over months rather than days [102, 140].

The fifth direction is to protect the idle moment, treating unfilled time as something to preserve rather than a gap to fill, through interfaces that end, feeds that stop, and systems that stay quiet when nothing needs saying. Boredom research suggests that unfilled time can be a resource [38], and designs that remove newsfeeds, autoplay and notifications show that this can be built [76, 124, 150]. The sixth is to bring reflective qualities to screens and AI. Screens make up 42% of the corpus and score among the lowest, and no AI system has invitational form. Since these are the forms that shape most people's everyday thinking, it is here that reflective computing matters most (Section 7.5).

### 7.8 Evaluating reflective computing

A paradigm needs ways of telling whether a system achieves its aims. Reviews have found that HCI often claims to support reflection without defining or evaluating it [7], and the corpus confirms that this remains common, as many systems report usage, satisfaction or behavioral outcomes rather than whether and how people reflected. Thus, reflective computing suggests evaluation at three levels.

At the design level, the codebook can be used to describe what a design expresses, before or after deployment. It is a profile, not a verdict, because two systems with the same proximity score can be reflective in very different ways. Therefore, profiles should be reported dimension by dimension. Its reliability (Section 5.1) indicates that form, openness, depth, span and tempo agency can be coded consistently by independent coders, while protected idleness and pace need the amended rules of v1.1.

At the experience level, whether people actually reflected is an empirical question that the rubric cannot answer. Existing frameworks offer starting points, namely the levels of reflection from description to transformation [46], the dimensions of breakdown, inquiry and transformation [8], and the design resources of temporal perspective, conversation, comparison and discovery [11]. Moreover, each dimension suggests its own questions. Did people feel in control of the pace? Did the form signal that pausing was welcome? Did they experience the system's quiet as rest or as absence? Did they trust that their reflections served them?

At the time level, evaluation must attend to time, because temporal span is itself a dimension. A reflective technology may have little visible effect in a short study and a significant one over months, as people return to it, lapse and come back. Hence, long field deployments of research products [140] and careful evaluation designs for everyday technologies [102] are better suited to this than laboratory studies.

### 7.9 A research agenda

The definition and dimensions are a proposal, and several questions follow from them. The first is whether the dimensions are complete and distinct. In this regard, they should be tested on new corpora and by other analysts. The blind open coding suggests three candidate gaps (contestable inference, automation that displaces noticing, and other-directed reflection), and social reflection, in which people reflect together [42, 75, 125, 137], did not emerge as a separate gap in Phase 1. Whether these are new dimensions or facets of existing ones is an empirical question for coding on new corpora, including a random sample of ACM publications. A second question is how the dimensions should be weighted. The proximity score treats them as equal and additive, whereas the factor structure suggests two composite scores, form and depth, with tempo agency loading moderately on depth. Whether some dimensions are preconditions for others remains open.

Three further questions concern design. What bridges the two cultures? Designs that combine invitational form with scaffolded depth are nearly absent from the corpus, with a single close call [79], and building and studying them would test whether the combination is more reflective than either alone. How can tempo agency coexist with designed slowness? The three approaches of Section 7.4 should be compared in deployment, to see which of them preserve the qualities of slow design while giving people control. What does reflective AI look like in practice? The moves outlined in Section 7.5 should be built and studied, particularly in writing, decision-making and learning, where generative AI is most widely used.

The last two questions concern context and cost. How does reflective computing work across cultures and contexts? The corpus is drawn from ACM venues dominated by Western, often domestic and workplace settings, and since practices of reflection, pace and idleness vary across cultures and communities, the dimensions may need to be adapted. What are the costs of reflection? Reflection is not always welcome. For people under stress or living with trauma, prompts to reflect may be unwelcome or harmful, and rumination is not reflection. Thus, it is imperative for reflective computing to understand when reflection helps and when it does not, which tempo agency and interpretive openness address partly but not fully.

### 7.10 Limitations

This study has seven main limitations.

First, the dimensions were derived and applied on the same corpus. The abductive analysis that produced the dimensions and the codebook that scored the systems both drew on the 134 systems. Hence, the Phase 2 profile describes how the corpus looks through the dimensions rather than independently confirming them. The blind second coder's open "other gap" notes provide a partial test, and they found no recurring absence of the same frequency as the seven dimensions. Nevertheless, a test on systems not used in the derivation is still required. To support such a test, we release the codebook, the coded corpus and the adjudication log.

Second, the exemplar corpus is purposive. It was assembled from expert knowledge, key reviews and citation chasing. As a result, paradigm counts reflect the corpus rather than the distribution of ACM publications, and the period analysis cannot separate time from paradigm. A random sample drawn from the ACM Digital Library would show whether the exemplar profile generalizes to typical publications.

Third, the coding rests on two coders, and the author adjudicated. The second coder was a human researcher, blind in the first round to the author's scores and throughout to the derivation of the dimensions, worked from a written codebook, and gave a reason for every score. However, the author both derived the dimensions and made every final decision, and most adjudicated disagreements were resolved toward the second coder's scores, largely through codebook amendments the author made after the first agreement round. Moreover, agreement between two coders cannot show how well the codebook transfers to analysts outside the project, and a third coder working on a stratified subset, that is, a subset chosen to include enough systems from each paradigm and score level, would test this.

Fourth, sources varied in the first round. The second coder first worked from the full text for 85 of 134 systems. All remaining papers were later supplied in full and re-coded, so the final scores rest on the full text for 134 systems. The first-round codes for those systems, and the changes the re-check made, are archived. Because this re-coding came after adjudication and used the amended codebook, it shows that the final codes hold under a full reading but is not an independent estimate of agreement. Hence, the reliability figures for independent coding are those of the first round.

Fifth, two codes needed amendment. Protected idleness and pace did not reach acceptable agreement under codebook v1.0. The amendments in v1.1 address the systematic sources of disagreement. However, their reliability should be re-estimated on new systems.

Sixth, the codebook scores what a design expresses as described in its paper, not how people experienced it in use. Thus, a system with a modest score may still have prompted deep reflection among its users, and a high-scoring system may not have.

Seventh, the proximity score treats the seven dimensions as equal and additive. The factor structure shows that they are not independent, and a system might be highly reflective through depth alone. Hence, the score is best analyzed as a summary of how many reflective qualities a design expresses, not as a measure of how reflective it is.

### 7.11 Implications for design, research and education

For designers, the dimensions offer a vocabulary for asking what a design does for reflection, and Table 11 offers a starting point within each paradigm, that is, to identify its characteristic gap and apply its translation move. Designers of AI systems can use the moves in Section 7.5 to ask whether their interfaces leave people the time and space to think. For reviewers and program committees, the codebook offers questions to ask of systems that claim to support reflection, namely who controls the pace, what the form signals, whether meaning is left open, how far reflection goes, over what period, what happens in idle moments, and whom the person's reflections serve.

For researchers, the definition and dimensions are a proposal to be tested. The archived corpus, codebook and adjudication log can be extended, re-coded or challenged, and the dimensions can be applied to new systems to see whether they hold. Finally, for educators and practitioners, the analysis suggests that reflection is not a feature to be added to a product but a set of qualities that run through pace, form, content, time, silence and ethics. Thus, a reflective prompt in an engagement-driven interface, or a slow artifact that cannot be paused, captures only part of what reflection requires.

## 8 CONCLUSION

This article defined reflective computing by reading thirty years of designed interventions for what they lack. Across 134 systems from 17 ACM venues, seven gaps recur, namely pace imposed by the system, form that does not invite pausing, verdicts that close down meaning, reflection that stops at description, momentary use, idle moments left unprotected, and inner states that are covertly influenced, exposed or left unaccounted for. Restated positively, these gaps define seven dimensions, namely tempo agency, invitational form, interpretive openness, reflective depth, temporal span, protected idleness and ethical non-extraction. Two independent coders applied five of them with at least tentative reliability, non-extraction with high agreement masked by the prevalence paradox, and protected idleness only after a codebook amendment, and the gaps persist among systems built for reflection.

Most dimensions have working exemplars. However, no system holds all seven. Slow and experience-centered designs show how form can invite pause and interpretation. Personal informatics and positive computing show how prompts and data can scaffold deeper thought. Connectedness and digital-wellbeing systems show that people value control over the pace of their technology, and across the corpus, designs that give it score higher on three of the six other dimensions, namely openness, depth and idleness. Yet the two cultures of reflection, through form and through scaffolded thought, have met in only one design. Thus, reflective computing names the whole these parts imply, that is, technology that gives people the time, the form and the questions to think for themselves, at a pace they choose.

### DECLARATIONS

**Data availability.** The full coding of all 134 systems, with a reason for every score, the codebook (v1.0 and v1.1), both coders' independent scores, the adjudication log and the analysis scripts are archived on OSF (anonymized view-only link for review: https://osf.io/yrxmb/overview?view_only=a44e9889d35a4c6f93e9bf5250b8e39b). Appendix A lists the consensus scores.

**Use of generative AI.** A large language model was used to edit the manuscript text sparsely. All coding and adjudication were carried out by the two human coders, the author and the second coder; no generative AI tool was used to code systems, prepare adjudication notes or decide scores. The author checked all such text against the data and sources and takes full responsibility for the content of the article.

**Disclosure of interest.** The author reports no conflict of interest.

**Funding.** This work was supported by the author's university seed grant.

## A CODED CORPUS

Consensus scores after independent double coding and adjudication. D1–D7: 0 absent, 1 partial, 2 strong; NR = paper silent on data use (scored as 1). Score = sum (0–14). Aim = reflection as a stated aim (2 primary, 1 secondary, 0 none). Src = the second coder's first-round source (F full text, an abstract, S secondary, M prior knowledge); systems not marked F were re-coded from the full text before final scores were set. A one-line reason for every score, both coders' independent scores and the adjudication rationale for every disputed cell is in Supplement S1.

Table A1: All 134 coded systems in chronological order (consensus scores).

| System [ref] | Venue | Paradigm | Pace | Form | D1 | D2 | D3 | D4 | D5 | D6 | D7 | Sum | Aim | Src |
|---|---|---|---|---|---|---|---|---|---|---|---|---|---|---|
| Audio Aura [134] | CHI 1998 | Calm | System | wearable | 0 | 1 | 1 | 0 | 0 | 1 | 1 | 4 | 0 | F |
| Water Lamp and Pinwheels [32] | CHI 1998 | Calm | System | ambient | 0 | 2 | 1 | 0 | 0 | 1 | NR | 5 | 0 | S |
| ambientROOM [87] | CHI 1998 | Calm | System | ambient | 1 | 2 | 1 | 0 | 1 | 1 | NR | 7 | 0 | S |
| inTouch [13] | CSCW 1998 | Connect. | User | tangible | 2 | 2 | 2 | 0 | 0 | 1 | NR | 8 | 0 | F |
| Babble [42] | CHI 1999 | Connect. | User | screen | 2 | 0 | 1 | 0 | 1 | 1 | 1 | 6 | 0 | F |
| Information Percolator [71] | UIST 1999 | Calm | System | ambient | 0 | 2 | 1 | 0 | 0 | 1 | NR | 5 | 0 | S |
| Casablanca [73] | CHI 2001 | Connect. | User | hybrid | 2 | 2 | 1 | 0 | 1 | 1 | 2 | 9 | 0 | A |
| Digital Family Portrait [135] | CHI 2001 | Calm | System | ambient | 0 | 2 | 1 | 0 | 1 | 1 | 1 | 6 | 1 | F |
| LumiTouch [18] | CHI EA 2001 | Affective | Shared | tangible | 1 | 2 | 2 | 0 | 0 | 1 | NR | 7 | 0 | F |
| Scope [173] | AVI 2002 | Attention | Shared | screen | 1 | 1 | 0 | 0 | 0 | 1 | NR | 4 | 0 | F |
| Sideshow [15] | CSCW 2002 | Attention | System | screen | 1 | 1 | 0 | 0 | 0 | 0 | NR | 3 | 0 | F |
| Drift Table [53] | CHI EA 2004 | Slow | Shared | tangible | 1 | 2 | 2 | 1 | 0 | 2 | NR | 9 | 1 | F |
| Lilsys [9] | CSCW 2004 | Attention | System | hybrid | 1 | 0 | 0 | 0 | 0 | 1 | 1 | 3 | 0 | F |
| Second Messenger [37] | CSCW 2004 | Persuasive | System | ambient | 0 | 0 | 0 | 1 | 0 | 1 | 0 | 2 | 1 | F |
| Breakaway [88] | CHI EA 2005 | Persuasive | System | ambient | 0 | 2 | 1 | 0 | 0 | 1 | 1 | 5 | 0 | A |
| Power-Aware Cord [66] | CHI EA 2005 | Persuasive | Shared | ambient | 1 | 2 | 1 | 1 | 0 | 1 | NR | 7 | 1 | S |
| Waterbot [3] | CHI 2005 | Persuasive | System | ambient | 0 | 1 | 0 | 0 | 1 | 0 | 0 | 2 | 0 | A |
| eMoto [168] | CHI EA 2005 | Affective | User | hybrid | 2 | 1 | 2 | 0 | 0 | 1 | NR | 7 | 0 | F |
| Affective Diary [117] | CHI EA 2006 | Affective | User | hybrid | 2 | 2 | 2 | 1 | 1 | 1 | NR | 10 | 2 | F |
| History Tablecloth [54] | DIS 2006 | Slow | Shared | ambient | 1 | 2 | 2 | 1 | 0 | 1 | NR | 8 | 2 | S |
| Houston [25] | CHI 2006 | Persuasive | Shared | screen | 1 | 0 | 0 | 0 | 1 | 0 | 1 | 3 | 1 | S |
| Sensual Evaluation Instrument [86] | CHI 2006 | Affective | User | tangible | 2 | 2 | 2 | 0 | 0 | 1 | NR | 8 | 0 | M |
| Home Health Horoscope [55] | CHI 2007 | Slow | System | ambient | 0 | 1 | 2 | 1 | 0 | 1 | 1 | 6 | 2 | F |
| EatWell [65] | CSCW 2008 | Connect. | User | convers. | 2 | 0 | 1 | 1 | 1 | 0 | 1 | 6 | 1 | F |
| MAHI [125] | CHI 2008 | Pers. inf. | Shared | hybrid | 1 | 0 | 1 | 2 | 1 | 1 | 1 | 7 | 2 | F |
| Meeting Mediator [95] | CSCW 2008 | Persuasive | System | wearable | 0 | 1 | 0 | 0 | 0 | 1 | 1 | 3 | 0 | F |

| System [ref] | Venue | Paradigm | Pace | Form | D1 | D2 | D3 | D4 | D5 | D6 | D7 | Sum | Aim | Src |
|---|---|---|---|---|---|---|---|---|---|---|---|---|---|---|
| Threshold Devices [56] | CHI 2008 | Slow | System | ambient | 0 | 2 | 2 | 1 | 0 | 1 | NR | 7 | 1 | F |
| UbiFit Garden [26] | CHI 2008 | Persuasive | Shared | screen | 1 | 1 | 1 | 1 | 1 | 0 | 1 | 6 | 1 | F |
| Baby Steps [94] | CHI 2009 | Pers. inf. | Shared | screen | 1 | 1 | 1 | 1 | 2 | 0 | 1 | 7 | 1 | S |
| Playful Bottle [20] | UbiComp 2009 | Persuasive | System | hybrid | 0 | 1 | 0 | 0 | 0 | 0 | 0 | 1 | 0 | A |
| Time capsule [148] | CHI 2009 | Slow | User | tangible | 2 | 1 | 2 | 1 | 2 | 1 | 1 | 10 | 2 | F |
| UbiGreen [49] | CHI 2009 | Persuasive | System | screen | 0 | 1 | 0 | 0 | 0 | 0 | 1 | 2 | 1 | F |
| Ambient influence [157] | UbiComp 2010 | Persuasive | System | ambient | 0 | 2 | 1 | 1 | 1 | 1 | NR | 7 | 1 | A |
| Family Archive [101] | CSCW 2010 | Slow | User | tangible | 2 | 1 | 1 | 1 | 1 | 1 | 1 | 8 | 1 | F |
| Mind the Body [161] | DIS 2010 | Affective | User | hybrid | 1 | 2 | 2 | 1 | 1 | 1 | NR | 9 | 2 | F |
| Prayer Companion [57] | CHI 2010 | Slow | System | ambient | 0 | 2 | 2 | 1 | 0 | 1 | NR | 7 | 1 | F |
| UpStream [111] | CHI 2010 | Persuasive | System | ambient | 0 | 1 | 0 | 0 | 0 | 1 | 1 | 3 | 1 | M |
| Fit4life [154] | CHI 2011 | Persuasive | System | wearable | 0 | 0 | 0 | 0 | 0 | 0 | 0 | 0 | 0 | A |
| Peripheral Paced Respiration [133] | UIST 2011 | Mindful. | System | screen | 0 | 1 | 0 | 0 | 0 | 0 | NR | 2 | 0 | F |
| Photostroller [58] | CHI 2011 | Slow | Shared | hybrid | 1 | 2 | 2 | 1 | 0 | 1 | NR | 8 | 1 | F |
| PlateMate [136] | UIST 2011 | Pers. inf. | User | screen | 1 | 0 | 0 | 0 | 0 | 1 | 1 | 3 | 1 | F |
| AffectAura [128] | CHI 2012 | Affective | User | screen | 2 | 1 | 0 | 1 | 1 | 1 | NR | 7 | 2 | F |
| BinCam [171] | CHI 2012 | Persuasive | Shared | hybrid | 1 | 0 | 1 | 1 | 1 | 1 | 0 | 5 | 1 | S |
| Email cut-off [126] | CHI 2012 | Attention | User | screen | 1 | 0 | 0 | 0 | 0 | 2 | NR | 4 | 0 | F |
| Fixture-level eco-feedback [50] | CHI 2012 | Persuasive | Shared | screen | 1 | 0 | 1 | 1 | 1 | 1 | 1 | 6 | 1 | F |
| Heart rate sharing [165] | CHI 2012 | Affective | User | screen | 1 | 1 | 1 | 0 | 0 | 1 | 1 | 5 | 0 | A |
| Lullaby [92] | UbiComp 2012 | Pers. inf. | User | hybrid | 1 | 0 | 1 | 1 | 1 | 1 | 1 | 6 | 1 | F |
| Mood Meter [72] | UbiComp 2012 | Affective | System | screen | 0 | 0 | 0 | 0 | 1 | 1 | 0 | 2 | 0 | F |
| Photobox [138] | DIS 2012 | Slow | System | tangible | 0 | 2 | 1 | 1 | 2 | 2 | NR | 9 | 1 | F |
| ShutEye [6] | CHI 2012 | Persuasive | System | screen | 1 | 2 | 0 | 0 | 0 | 1 | 1 | 5 | 1 | S |
| Sonic Cradle [175] | DIS 2012 | Mindful. | User | immersive | 2 | 2 | 2 | 1 | 0 | 2 | NR | 10 | 1 | S |
| StressSense [119] | UbiComp 2012 | Affective | System | screen | 0 | 0 | 0 | 0 | 0 | 1 | 1 | 2 | 1 | F |
| Dubuque Electricity Portal [43] | CHI 2013 | Persuasive | User | screen | 2 | 0 | 1 | 0 | 1 | 1 | 1 | 6 | 1 | A |
| Echo [85] | CHI 2013 | Positive | Shared | screen | 1 | 1 | 1 | 2 | 2 | 0 | 1 | 8 | 2 | A |
| Health Mashups [10] | TOCHI 2013 | Pers. inf. | Shared | screen | 1 | 0 | 1 | 1 | 2 | 0 | NR | 6 | 1 | S |
| Indoor Weather Stations [59] | CHI 2013 | Slow | Shared | ambient | 1 | 2 | 2 | 1 | 1 | 1 | 1 | 9 | 1 | F |
| Personhood jewelry [176] | CHI 2013 | Slow | User | wearable | 2 | 2 | 2 | 1 | 2 | 1 | NR | 11 | 2 | S |
| Reconsidered bathroom scale [93] | UbiComp 2013 | Pers. inf. | User | tangible | 2 | 1 | 1 | 1 | 1 | 1 | NR | 8 | 1 | F |
| ShareTable [181] | CSCW 2013 | Connect. | User | hybrid | 2 | 1 | 1 | 0 | 0 | 0 | 1 | 5 | 0 | A |
| Spheres of Wellbeing [172] | CHI 2013 | Mindful. | User | tangible | 2 | 2 | 1 | 1 | 1 | 1 | 2 | 10 | 1 | S |

| System [ref] | Venue | Paradigm | Pace | Form | D1 | D2 | D3 | D4 | D5 | D6 | D7 | Sum | Aim | Src |
|---|---|---|---|---|---|---|---|---|---|---|---|---|---|---|
| Counterfunctional things [151] | DIS 2014 | Slow | User | tangible | 1 | 2 | 1 | 1 | 1 | 2 | NR | 9 | 0 | F |
| Photobox field study [139] | CHI 2014 | Slow | System | tangible | 0 | 2 | 2 | 1 | 2 | 2 | NR | 10 | 1 | F |
| Visual cuts [40] | DIS 2014 | Pers. inf. | User | screen | 1 | 0 | 1 | 2 | 1 | 1 | 1 | 7 | 2 | F |
| Adaptive stress reduction [108] | CHI 2015 | Positive | System | screen | 0 | 0 | 0 | 0 | 1 | 0 | NR | 2 | 0 | A |
| EmotiCal [77] | CHI 2015 | Positive | Shared | screen | 1 | 0 | 1 | 1 | 1 | 1 | 1 | 6 | 1 | M |
| Energy Babble [60] | CHI 2015 | Slow | System | ambient | 0 | 2 | 2 | 1 | 0 | 0 | 1 | 6 | 1 | F |
| FamiLync [103] | UbiComp 2015 | Dig. wellb. | User | screen | 1 | 0 | 1 | 1 | 1 | 1 | 1 | 6 | 1 | F |
| MoodLight [166] | CSCW 2015 | Affective | System | ambient | 1 | 2 | 1 | 1 | 0 | 1 | 1 | 7 | 1 | F |
| MyBehavior [155] | UbiComp 2015 | Persuasive | Shared | screen | 1 | 0 | 0 | 0 | 0 | 1 | 1 | 3 | 0 | F |
| NUGU [104] | CSCW 2015 | Dig. wellb. | Shared | screen | 1 | 0 | 0 | 1 | 1 | 2 | 1 | 6 | 1 | F |
| SleepTight [21] | UbiComp 2015 | Pers. inf. | User | screen | 2 | 0 | 1 | 1 | 1 | 1 | 1 | 7 | 2 | F |
| Spaceship Launch [160] | CSCW 2015 | Persuasive | User | screen | 1 | 1 | 0 | 1 | 1 | 1 | NR | 6 | 1 | A |
| Datacatcher [61] | CHI 2016 | Slow | Shared | tangible | 1 | 1 | 2 | 1 | 1 | 1 | NR | 8 | 1 | A |
| EmotionCheck [28] | UbiComp 2016 | Affective | System | wearable | 1 | 1 | 0 | 0 | 0 | 1 | 0 | 3 | 0 | S |
| Hint (color-changing shirt) [83] | DIS 2016 | Affective | System | wearable | 0 | 2 | 2 | 1 | 0 | 1 | 0 | 6 | 1 | F |
| Lock n' LoL [105] | CHI 2016 | Dig. wellb. | Shared | screen | 1 | 0 | 0 | 0 | 1 | 2 | 1 | 5 | 0 | F |
| MyTime [74] | CHI 2016 | Dig. wellb. | Shared | screen | 1 | 0 | 0 | 1 | 0 | 0 | NR | 3 | 1 | F |
| Physikit [82] | CHI 2016 | Calm | Shared | ambient | 1 | 2 | 1 | 1 | 0 | 1 | 1 | 7 | 1 | F |
| RelaWorld [109] | IUI 2016 | Mindful. | System | immersive | 0 | 1 | 0 | 0 | 0 | 1 | NR | 3 | 0 | A |
| SleepCoacher [35] | UIST 2016 | Pers. inf. | System | hybrid | 1 | 0 | 0 | 1 | 1 | 0 | 1 | 4 | 0 | F |
| Somaesthetic appreciation design [79] | CHI 2016 | Mindful. | Shared | immersive | 1 | 2 | 2 | 2 | 0 | 2 | NR | 10 | 2 | S |
| TimeAware [96] | CHI 2016 | Dig. wellb. | Shared | screen | 1 | 0 | 0 | 1 | 1 | 1 | NR | 5 | 1 | F |
| Unlock Journaling [183] | CHI 2016 | Pers. inf. | User | screen | 2 | 0 | 1 | 1 | 1 | 0 | NR | 6 | 1 | M |
| 24 hours without notifications [150] | MobileHCI 2017 | Dig. wellb. | User | screen | 2 | 0 | 0 | 1 | 0 | 2 | NR | 6 | 0 | F |
| BreakSense [17] | CHI 2017 | Persuasive | Shared | hybrid | 1 | 0 | 0 | 0 | 0 | 0 | 1 | 2 | 0 | F |
| BrightBeat [62] | CHI EA 2017 | Mindful. | System | ambient | 1 | 2 | 0 | 0 | 0 | 1 | 0 | 4 | 0 | A |
| EngageMeter [69] | CHI 2017 | Affective | System | hybrid | 1 | 0 | 0 | 0 | 0 | 1 | 0 | 2 | 1 | F |
| FlowLight [186] | CHI 2017 | Attention | System | ambient | 1 | 1 | 0 | 0 | 0 | 1 | 1 | 4 | 0 | S |
| HeartChat [70] | CHI 2017 | Affective | Shared | screen | 1 | 0 | 1 | 0 | 1 | 1 | 1 | 5 | 1 | F |
| Inner Garden [158] | CHI 2017 | Mindful. | User | tangible | 2 | 2 | 1 | 1 | 2 | 2 | 1 | 11 | 1 | S |
| Let's FOCUS [97] | IMWUT 2017 | Dig. wellb. | Shared | screen | 1 | 0 | 0 | 1 | 1 | 1 | 0 | 4 | 1 | F |
| Life Tree [146] | CHI PLAY 2017 | Mindful. | Shared | immersive | 1 | 2 | 0 | 1 | 0 | 1 | NR | 6 | 1 | F |
| OmniTrack [98] | IMWUT 2017 | Pers. inf. | User | screen | 2 | 0 | 1 | 1 | 1 | 1 | 1 | 7 | 1 | F |

| System [ref] | Venue | Paradigm | Pace | Form | D1 | D2 | D3 | D4 | D5 | D6 | D7 | Sum | Aim | Src |
|---|---|---|---|---|---|---|---|---|---|---|---|---|---|---|
| Plan & Play [75] | IDC 2017 | Dig. wellb. | Shared | screen | 1 | 0 | 0 | 1 | 0 | 1 | NR | 4 | 1 | F |
| Self-experimentation approaches [112] | CHI 2017 | Pers. inf. | Shared | hybrid | 1 | 0 | 1 | 2 | 1 | 0 | NR | 6 | 1 | S |
| TummyTrials [91] | CHI 2017 | Pers. inf. | Shared | screen | 1 | 0 | 0 | 2 | 1 | 0 | NR | 5 | 1 | F |
| WorkAnalytics [129] | CSCW 2017 | Pers. inf. | Shared | screen | 1 | 0 | 1 | 2 | 1 | 0 | 1 | 6 | 1 | S |
| Attending to Breath [153] | DIS 2018 | Mindful. | Shared | immersive | 1 | 2 | 2 | 0 | 0 | 2 | NR | 8 | 0 | F |
| Breeze [48] | CHI 2018 | Affective | System | wearable | 0 | 2 | 1 | 0 | 0 | 1 | 1 | 5 | 0 | F |
| Coco's Videos [76] | CHI 2018 | Dig. wellb. | Shared | screen | 1 | 0 | 1 | 0 | 0 | 2 | 1 | 5 | 1 | F |
| Just Breathe [145] | IMWUT 2018 | Mindful. | System | hybrid | 0 | 1 | 0 | 0 | 0 | 1 | NR | 3 | 0 | F |
| Olly and Slow Game [141] | CHI 2018 | Slow | Shared | tangible | 1 | 2 | 2 | 1 | 2 | 2 | NR | 11 | 1 | S |
| Peer support chats [137] | CHI 2018 | Positive | System | convers. | 1 | 0 | 1 | 2 | 0 | 0 | 1 | 5 | 1 | F |
| Pocket Skills [163] | CHI 2018 | Positive | Shared | convers. | 1 | 0 | 0 | 1 | 1 | 0 | 1 | 4 | 1 | M |
| Reflection Companion [107] | IMWUT 2018 | Pers. inf. | Shared | convers. | 1 | 0 | 2 | 2 | 1 | 0 | NR | 7 | 2 | F |
| Ripple (emotional biosensing) [84] | CSCW 2018 | Affective | System | wearable | 0 | 2 | 2 | 1 | 0 | 1 | 1 | 7 | 1 | S |
| Robota [106] | DIS 2018 | Pers. inf. | System | convers. | 1 | 0 | 2 | 2 | 1 | 0 | 1 | 7 | 2 | A |
| Time for Break [122] | CHI 2018 | Persuasive | System | screen | 1 | 0 | 0 | 0 | 0 | 0 | 1 | 2 | 0 | F |
| Vibration nudge [143] | MobileHCI 2018 | Dig. wellb. | System | screen | 0 | 0 | 0 | 0 | 0 | 0 | 1 | 1 | 1 | A |
| Voice self-perception [29] | CHI 2018 | Affective | System | wearable | 0 | 0 | 0 | 0 | 0 | 1 | 0 | 1 | 0 | F |
| Workplace detachment agent [180] | CHI 2018 | Attention | User | convers. | 2 | 0 | 2 | 1 | 1 | 1 | 1 | 8 | 1 | F |
| Animo [118] | IMWUT 2019 | Affective | Shared | wearable | 1 | 1 | 2 | 0 | 0 | 0 | 1 | 5 | 0 | F |
| BoostMeUp [30] | IMWUT 2019 | Affective | System | wearable | 1 | 1 | 0 | 0 | 0 | 1 | 1 | 4 | 0 | A |
| Chronoscope [19] | DIS 2019 | Slow | User | tangible | 2 | 2 | 2 | 1 | 2 | 1 | NR | 11 | 2 | M |
| GoalKeeper [99] | IMWUT 2019 | Dig. wellb. | System | screen | 1 | 0 | 0 | 1 | 1 | 0 | NR | 4 | 1 | F |
| Olly field study [142] | CHI 2019 | Slow | System | tangible | 1 | 2 | 2 | 1 | 2 | 2 | 1 | 11 | 2 | A |
| Balance, Cogito and Dott [131] | TEI 2020 | Slow | User | tangible | 2 | 2 | 2 | 1 | 1 | 1 | 1 | 10 | 2 | F |
| Confidence and explanation display [184] | FAT* 2020 | Human-AI | System | screen | 0 | 0 | 0 | 0 | 0 | 1 | NR | 2 | 0 | F |
| Facebook self-control interventions [124] | CHI 2020 | Dig. wellb. | Shared | screen | 1 | 0 | 1 | 1 | 0 | 2 | 1 | 6 | 1 | F |
| Mirror Ritual [156] | CHI 2020 | Affective | User | ambient | 2 | 2 | 2 | 1 | 0 | 2 | NR | 10 | 2 | F |
| PIV breathing pacer [130] | CHI 2020 | Mindful. | System | wearable | 0 | 1 | 0 | 0 | 0 | 1 | NR | 3 | 0 | S |
| Trackly [4] | CHI 2020 | Pers. inf. | User | screen | 2 | 2 | 1 | 1 | 1 | 1 | NR | 9 | 2 | F |
| AI explanations in teams [5] | CHI 2021 | Human-AI | Shared | screen | 1 | 0 | 0 | 0 | 0 | 1 | NR | 3 | 0 | F |
| Cognitive forcing functions [14] | CSCW 2021 | Human-AI | Shared | screen | 1 | 0 | 0 | 1 | 0 | 1 | NR | 4 | 1 | F |
| AI assistance and incidental learning [51] | IUI 2022 | Human-AI | System | screen | 0 | 0 | 1 | 1 | 0 | 1 | NR | 4 | 0 | F |
| CoAuthor [113] | CHI 2022 | Human-AI | User | screen | 2 | 0 | 0 | 0 | 0 | 1 | 1 | 4 | 0 | F |

| System [ref] | Venue | Paradigm | Pace | Form | D1 | D2 | D3 | D4 | D5 | D6 | D7 | Sum | Aim | Src |
|---|---|---|---|---|---|---|---|---|---|---|---|---|---|---|
| Continuous text summaries [33] | UIST 2022 | Human-AI | User | screen | 2 | 0 | 1 | 1 | 0 | 1 | 1 | 6 | 2 | F |
| Wordcraft [182] | IUI 2022 | Human-AI | User | screen | 2 | 0 | 0 | 0 | 0 | 1 | 1 | 4 | 0 | S |
| CatAlyst [2] | CHI 2023 | Human-AI | System | screen | 1 | 0 | 0 | 0 | 0 | 0 | 1 | 2 | 0 | F |
| Explanations and overreliance [174] | CSCW 2023 | Human-AI | System | screen | 0 | 0 | 0 | 0 | 0 | 1 | NR | 2 | 0 | F |
| Opinionated LLM co-writing [89] | CHI 2023 | Human-AI | Shared | screen | 1 | 0 | 0 | 0 | 0 | 0 | 0 | 1 | 0 | F |
| Question-framed explanations [34] | CHI 2023 | Human-AI | User | screen | 1 | 0 | 2 | 2 | 0 | 1 | NR | 7 | 2 | A |
| SwitchTube [121] | CHI 2023 | Dig. wellb. | User | screen | 2 | 0 | 1 | 1 | 0 | 2 | 1 | 7 | 0 | F |
| MindfulDiary [100] | CHI 2024 | Human-AI | User | convers. | 2 | 0 | 1 | 2 | 1 | 1 | 1 | 8 | 2 | F |